\documentclass[twoside,leqno]{article}

\usepackage[letterpaper]{geometry}
\usepackage[linesnumbered,ruled]{algorithm2e}
\usepackage{amsmath,amsfonts,amssymb}
\usepackage[colorlinks=true, allcolors=blue]{hyperref}
\usepackage[nameinlink,capitalise]{cleveref}

\crefalias{customsection}{section}
\crefname{customsection}{\S}{\S}
\Crefname{customsection}{\S}{\S}
\creflabelformat{customsection}{#2\textup{#1}#3}

\usepackage{ltexpprt}
\pdfoutput=1

\usepackage[utf8]{inputenc}

\usepackage{latexsym}
\usepackage[noadjust]{cite}
\usepackage{dirtytalk}
\usepackage{mathtools}
\usepackage{thm-restate}
\usepackage{enumitem}

\newcommand{\F}{\mathbb{F}}

\newcommand{\E}{\mathbb{E}}
\newcommand{\calD}{\mathcal{D}}
\newcommand{\calC}{\mathcal{C}}

\newcommand{\calP}{\mathcal{P}}

\newcommand{\zo}{\{0, 1\}}

\newcommand{\Density}{\text{Density}}

\newcommand{\Zivny}{Zivn\'y}

\newcommand{\eps}{\varepsilon}

\newcommand{\ra}{\rightarrow}

\newcommand{\wt}{\mathrm{wt}}

\newcommand{\AND}{\wedge}
\newcommand{\Span}{\text{Span}}

\newcommand{\Supp}{\mathrm{Supp}}

\newcommand{\Deven}{\calD_{\text{even}}}
\newcommand{\Dodd}{\calD_{\text{odd}}}
\newcommand{\teveni}{k_{\text{even}, i}}

\title{Code Sparsification and its Applications}
\author{Sanjeev Khanna\thanks{School of Engineering and Applied Sciences, University of Pennsylvania, Philadelphia, PA. Email: {\tt sanjeev@cis.upenn.edu}. Supported in part by NSF awards CCF-1934876 and CCF-2008305.} \and Aaron (Louie) Putterman\thanks{School of Engineering and Applied Sciences, Harvard University, Cambridge, Massachusetts, USA. Supported in party by the Simons Investigator Fellowship of Boaz Barak, NSF grant DMS-2134157, DARPA grant W911NF2010021, and DOE grant DE-SC0022199. Supported in part by the Simons Investigator Award of Madhu Sudan and NSF Award CCF 2152413. Email: \texttt{aputterman@g.harvard.edu}.} \and Madhu Sudan\thanks{School of Engineering and Applied Sciences, Harvard University, Cambridge, Massachusetts, USA. Supported in part by a Simons Investigator Award and NSF Award CCF 2152413. Email: \texttt{madhu@cs.harvard.edu}.}}

\date{\today}

\begin{document}

\maketitle


\begin{abstract}
  
We introduce a notion of \emph{code} sparsification that generalizes the notion of cut sparsification in graphs. 
For a (linear) code $\mathcal{C} \subseteq \mathbb{F}_q^n$ of dimension $k$ a $(1 \pm \epsilon)$-\emph{sparsification} of size $s$ is given by a weighted set  $S \subseteq [n]$ with $|S| \leq s$ such that for every codeword $c \in \mathcal{C}$ the projection $c|_S$ of $c$ to the set $S$ has (weighted) hamming weight which is a $(1 \pm \epsilon)$ approximation of the hamming weight of $c$. 
We show that for every code there exists a $(1 \pm \epsilon)$-sparsification of size $s = \widetilde{O}(k \log (q) / \epsilon^2)$. This immediately implies known results on graph and hypergraph cut sparsification up to polylogarithmic factors (with a simple unified proof) --- the former follows from the well-known fact that cuts in a graph form a linear code over $\mathbb{F}_2$, while the latter is obtained by a simple encoding of hypergraph cuts. 
Further, by connections between the eigenvalues of the Laplacians of Cayley graphs over $\mathbb{F}_2^k$ to the weights of codewords, we also give the first proof of the existence of spectral Cayley graph sparsifiers over $\mathbb{F}_2^k$ by Cayley graphs, i.e., where we sparsify the set of generators to nearly-optimal size. 
Additionally, this work can be viewed as a continuation of a line of works on building sparsifiers for constraint satisfaction problems (CSPs); this result shows that there exist near-linear size sparsifiers for CSPs over $\mathbb{F}_p$-valued variables whose unsatisfying assignments can be expressed as the zeros of a linear equation modulo a prime $p$.
As an application we give a full characterization of ternary Boolean CSPs (CSPs where the underlying predicate acts on three Boolean variables) that allow for near-linear size sparsification.
This makes progress on a question posed by Kogan and Krauthgamer (ITCS 2015) asking which CSPs allow for near-linear size sparsifiers (in the number of variables). 
    
At the heart  of our result is a codeword counting bound that we believe is of independent interest. Indeed, extending Karger's cut-counting bound (SODA 1993), we show a novel decomposition theorem of linear codes: we show that every linear code has a (relatively) small subset of coordinates such that after deleting 
those coordinates, the code on the remaining coordinates has a smooth upper bound on the number of codewords of small weight. Using the deleted coordinates in addition to a (weighted) random sample of the remaining coordinates now allows us to sparsify the whole code. The proof of this decomposition theorem extends Karger's proof (and the contraction method) in a clean way, while enabling the extensions listed above without any additional complexity in the proofs. 
\end{abstract}
\pagebreak
{\footnotesize \tableofcontents } 

\newpage 

\pagenumbering{arabic}

\section{Introduction}

In this work we introduce a notion of sparsification for linear codes, prove that ``nearly linear-size'' sparsifications exist for every linear code, and give some applications beyond coding theory. 
 We start by recalling the notion of sparsification and give some background.

Sparsification of data with respect to a certain class of queries alludes to a compression of the data that allows all queries in the class to be answered (or estimated) correctly. It has emerged as a fundamental concept in theoretical computer science, with graph sparsification for cut-queries being a central example. In seminal works, Karger~\cite{Kar94sparsification} and Bencz\'ur and Karger~\cite{BK96} showed how graphs may be sparsified by carefully sampling a subset of its edges and assigning weights to them,  so that for every cut, the cut-size in the sampled weighted graph gives a good estimate of the cut size in the input graph. Crucially the number of sampled edges was nearly linear-sized in the number of vertices of the graph. This work led to many strengthenings (in particular to allowing the sample to be linear sized~\cite{BSS09}), extensions (in particular to more general spectral notions of sparsification~\cite{ST11,BSS09}, to hypergraphs~\cite{KK15,CKN20, KKTY21a}), more recently to sparsifying sums of norms \cite{JLLS23}, and applications to a host of problems including solvers for max-flow and min-cut \cite{She13,KLOS14,Pen16}, as well as to better solvers for structured linear systems \cite{ST11,CKMPPRX14,KLPSS16,LS18,JS20} and applications to clustering \cite{CSWZ16}. In the streaming and sketching settings, small representations of graphs are equally important and work on graph sparsifiers has played a key role. 

A natural question that arises is which other classes of structures and queries allow for such sparsification. In this work we explore this question in a new terrain, namely linear codes, where the queries specify a message and the goal is to estimate the Hamming weight of its encoding under the linear code. We describe our setting formally first before motivating the problem. 

\subsection{Code Sparsification}

Throughout this paper $q$ will be a prime power and $\F_q$ will denote the finite field on $q$ elements. A linear code $\calC$ is a $\F_q$-linear subspace of $\F_q^n$, and we will assume it is the image of a linear map $E:\F_q^k \to \F_q^n$. The Hamming weight of a vector $v \in \F_q^n$, denoted $\wt(v)$ is the number of non-zero coordinates of $v$. Given a sequence of non-negative (integer) weights $w = (w_1,\ldots,w_n)$, the weighted Hamming weight of $v = (v_1,\ldots,v_n)$, denoted $\wt_w(v)$, equals $\sum_{i|v_i \ne 0} w_i$. For a vector $v\in \F_q^n$ and set $S \subseteq [n]$ the puncturing of $v$ to the set $S$, denoted $v|_S$ is the vector $(v_i)_{i \in S}$. The puncturing of a code $\calC \subseteq \F_q^n$ to the coordinates $S \subseteq [n]$ is the code $\calC|_S \subseteq \F_q^{|S|}$ given by $\calC|_S = \{v|_S :v \in \calC\}$. We are now ready to define code sparsifiers.

\begin{Definition}[Code Sparsifier]
For integer $s$, real $\epsilon > 0$ and a linear code $\calC \subseteq \F_q^n$, a $(1 \pm \epsilon)$-sparsifier of size $s$ for the code $\calC$ is a subset $S \subseteq [n]$ with $|S| \leq s$ along with weights $w_S = (w_i)_{i\in S}$ such that for every codeword $v \in \calC$, we have 
$$ (1-\epsilon) \wt(v) \leq \wt_w(v|_S) \leq (1+\epsilon) \wt(v).$$
(In other words the weighted Hamming weight of every codeword in the punctured code roughly equals its weight in the unpunctured code.) 
\end{Definition}

The vanilla representation of a linear code would involve $kn$ elements of $\F_q$. The sparsification reduces the representation size to $sk$ field elements which may be significantly smaller if $s \ll n$. In several applications we consider later, the code $\calC$ itself is obtained by puncturing a known fixed mother code $M \subseteq \F_q^N$. In such cases the vanilla representation of $\calC$ (to someone who knows $M$) would require $n \log N$ bits while the sparsification would require only $s \log N$ bits to describe. Thus in both cases the sparsification definitely compresses the representation of $\calC$. And if we fix any linear encoding scheme $E:\F_q^k \to\F_q^n$ such that $\calC = \{E(m) | m \in \F_q^k\}$, then the sparsification allows us to estimate the hamming weight of the encoding $E(m)$ of every message $m \in \F_q^k$. 
Thus, the definition of code sparsifiers fits the general notion of sparsification, and so we turn to the motivation for studying this concept. 

\subsection{Motivation} 

Our initial motivation for studying code sparsification is that it abstracts and generalizes cut sparsification in graphs. Specifically for every graph there is a linear code over $\F_2$ such that codewords of this code are indicator vectors of the edges crossing cuts in the graph. (This code is obtained by viewing the edge-vertex incidence matrix as the generator of the code.) Thus, a sparsifier for this code corresponds to a cut sparsifier for the associated graph. Existence of graph sparsifiers is typically proved by combinatorial or spectral analysis --- tools that are less amenable to application over codes. Thus the exploration of code sparsification forces us to revisit methods for constructing graph sparsifiers and extract the essential elements in this toolkit,

One broad class of sparsifiers that overlap significantly with code sparsifiers are {\em CSP sparsifiers}, introduced by Kogan and Krauthgamer~\cite{KK15} and studied further by Filtser and Krauthgamer~\cite{FK17} and Butti and \Zivny~\cite{BZ20}. Constraint Satisfaction Problems (CSPs) have as instances $n$ constraints on $k$ variables where each constraint operates on a constant number of variables (called the arity of the constraint). A CSP sparsifier aims to compress an instance of the CSP into a smaller (weighted) one on the same set of variables such that for every assignment to the variables, the sparsified CSP satisfies roughly the same number of constraints as the original one. When the variables take on values in a finite set $\F_q$ and the constraints are linear constraints over $\F_q$, then the sparsification task is a code sparsification task. (Note that code sparsification allows $q$ as well as the arity of the constraints to be non-constant and so code sparsification is not a subclass of CSP sparsification.)  In particular, cut sparsification is also a special case of CSP sparsification. 
Prior work had shown how to get nearly linear size sparsifiers for CSPs beyond cut sparsifiers. Specifically, in \cite{KK15}, it was shown that $r$-SAT instances on a universe of $k$ Boolean variables admit sparsifiers of size $\widetilde{O}(kr / \eps^2)$. This was improved to $\widetilde{O}(k / \eps^2)$ by Chen, Khanna and Nagda~\cite{CKN20}, but the family of CSPs for which this result holds was not broadened. The works \cite{FK17,BZ20} completely classify all binary CSPs (i.e., with arity two), that allow for nearly linear size sparsification, but a classification beyond $r=2$ remains wide open. Indeed \cite{FK17}, pose this as an open question, and \cite{BZ20} highlight the challenge of sparsifying $r$-XOR CSPs (for $r \geq 3$) as a central problem that remains unaddressed by graph and hypergraph sparsification techniques. Thus, code sparsification seems like the natural next frontier in CSP sparsification and worthy of further attention.

Finally we also give some new applications of code sparsification in this paper itself. In particular, we give a simple reduction from hypergraph cut sparsification to code sparsification that makes the former a special case of the latter (although there may be some polylogarithmic losses in the size of the sparsification). We also show how code sparsification can be used to derive structured sparsification of Cayley graphs over $\F_2^k$, by other Cayley graphs! We elaborate on these new connections and their implications after describing our main results.

\subsection{Main Results}

Our main theorem in this paper shows that every (possibly weighted) linear code $\calC \subseteq \F_q^n$ of dimension $k$ has a nearly-linear sized sparsifier, i.e., one of size $\widetilde{O}(\frac{k}{\epsilon^2}\log q)$.\footnote{In this paper we use the notation $\widetilde{O}(\cdot)$ to hide poly logarithmic factors in the argument.}

\begin{theorem}\label{thm:main}
    For every $\epsilon > 0$, prime power $q$, positive integers $k$ and $n$, every (possibly weighted) linear code $\calC \subseteq \F_q^n$ of dimension $k$ has a $\left(1 \pm \epsilon\right)$-code sparsifier of size $\widetilde{O}\left(\frac{k}{\epsilon^2}\log q\right)$. 
\end{theorem}

Note that the theorem is essentially optimal up to polylogarithmic factors in $k$ (for constant $\epsilon$ and $q$) in that codes require $\Omega(k)$-sized sparsifiers. 
Note also that  our result qualitatively reproduces the existential part of the sparsification result in \cite{BK96} while extending it vastly. Of particular interest is the fact that our result does not require the generator matrix of the code to be sparse, something that was evidently true of all previous works on sparsification, and potentially used as a core ingredient in many proofs. In fact, the theorem as stated above is completely independent of the choice of the generator matrix of the code, whereas previous analyses even for the graph-theoretic codes seem to rely on the use of a specific generator matrix. 

A central tool in the cut sparsifiers of \cite{Kar94sparsification,Kar99, BK96, FHH11} is ``Karger's cut counting bound''~\cite{Kar93, Kar99} which asserts that every graph on $k$ vertices whose minimum cut is $c$ has at most $k^{2\alpha}$ cuts of size at most $\alpha c$, for every integer $\alpha$. 
This bound can be interpreted in coding terms --- the minimum cut size is the minimum distance of the corresponding code, and distance is of course a central concept in coding theory. However the lemma is patently false for general codes. Specifically for codes of minimum distance $\Omega(n)$ the bound would suggest that there are at most $n^{O(1)}$ codewords in the code and every asymptotically good code (those which $k = \Omega(n)$) are counterexamples to this potential extension. In view of the centrality of this bound though, it is natural to ask what weaker bound one can get for general codes. Our next theorem gives a simple weakening that essentially suggests that the only counterexamples come from ``good'' codes embedded in $\calC$.  

\begin{theorem}\label{thm:newKargerIntro}
    For every  prime power $q$, parameters $d$, $k$ and $n$ and every linear code $\calC \subseteq \F_q^n$, the following holds: there exists a subset $T \subseteq [n]$ with $|T| \leq k \cdot d$ such that for $S = [n]\setminus T$ the code $\calC|_S$ satisfies the condition that for every integer $\alpha \geq 1$ the code $\calC|_S$ has at most $q^{\alpha}\cdot \binom{k}\alpha$ codewords of weight at most $\alpha d$. 
\end{theorem}

Note that in the above theorem, $d$ does not refer to the distance, but rather is a parameter of our choosing.

In other words while $\calC$ may have many relatively small weight codewords, they come from a sub-code $\calC|_T$ contained on a small set of coordinates while the rest of the code $\calC|_S$ has a smooth growth in the number of codewords of small weight. We note that this basic theorem about linear spaces does not seem to have been noticed before and could be of independent interest.

While it is immediate that \Cref{thm:newKargerIntro} can be used to get some sparsification for some codes, it is not clear how to use it to go all the way to \Cref{thm:main}. Indeed in the case of graph sparsification for preserving cuts, known proofs utilize additional notions from graph theory to identify importance of a coordinate (edge). For instance,~\cite{BK96} utilizes the notion of edge strengths while~\cite{FHH11} relies on edge connectivity to determine sampling probabilities, and in both cases, the analysis uses graph-theoretic structure to establish correctness of the resulting sparsifiers. 
We show nevertheless that a simple recursive scheme can be applied to sparsify every code. Indeed even the specialization of this proof to the graph-theoretic case of cut sparsifiers seems new and we describe this simpler proof in \Cref{sec:simplerBK}. 

We remark that one weakness of our results (or a major open question) is that our results are existential and we do not have efficient algorithms to produce the sparsifiers that we show exist. The difficulty roughly emerges from the difficulty of finding and counting low weight codewords in a code which are known hard problems in coding theory. 

\subsection{Applications}

\paragraph{Hypergraph Cut Sparsification.}

A cut sparsifier for a hypergraph is a simple extension of the notion of a cut sparsifier for graphs. Specifically it is a weighted subgraph of the input hypergraph such that for every 2-coloring of the vertices, the number of bichromatic edges in the original hypergraph is approximately the same as the weight of the bichromatic edges in the subgraph. Previous works by Kogan and Krauthgamer~\cite{KK15} (in the constant arity hyperedge case) and ultimately Chen, Khanna and Nagda~\cite{CKN20} (in the unbounded arity case) have given cut-sparsifiers of size $O(k \log (k) /\epsilon^2)$ for every hypergraph on $k$ vertices. We are able to recover their result qualitatively (up to polylogarithmic factors in $k$ and $1/\epsilon$) with a very simple reduction. (Specifically we note that if we choose $q$ to be a large enough prime and associate an $r$-vertex hyperedge with the vector $(q-r+1,1,\ldots,1,0,\ldots,0) \in \F_q^k$ then the only coordinates in encodings of messages in $\{0,1\}^k$ that are $0$ are the monochromatic edges.) See \Cref{rmk:hypergraphEncoding} for more details.  Indeed by applying this reduction to \Cref{thm:newKargerIntro} we also obtain a structural decomposition theorem for hypergraphs that does not seem to have been noticed before.

\begin{theorem}\label{thm:hypergraphIntro}
    For every integer $d\geq 1$, every hypergraph $H$ on $k$ vertices has a set of at most $kd$ hyperedges such that upon their removal, the resulting hypergraph satisfies the condition that for every integer $\alpha \geq 1$ it has at most $(2k)^{2\alpha}$ cuts of size $\leq \alpha d$.
\end{theorem}

Indeed, as in the setting of codes, an analog of \say{Karger's cut-counting bound} does not hold in the realm of hypergraphs \cite{KK15}. Thus, our analysis of code sparsifiers provides a more universal counting bound which decomposes codes and hypergraphs alike.

\paragraph{Cayley Graph Sparsifiers.}

A well-studied notion extending that of a cut-sparsifier for graphs is a spectral sparsifier. Formally a spectral sparsifier of a graph is a weighted subgraph whose Laplacian has eigenvalues close to that of Laplacian of the original graph. (The Laplacian of a graph $G = (V, E)$,  denoted $L_G$, is a $|V| \times |V|$ matrix, whose diagonal entries $L_{G_{i, i}}$ are the degrees of the $i$th vertex, and whose off diagonal entries $L_{G_{i, j}}$ are $-w_{i,j}$ where $w_{ij}$ is the weight of the edge $(i,j)$ in $G$.) Informally, a spectral sparsifier allows us to estimate the quadratic form $x^T L_G x$ for every real vector $x$, whereas a cut-sparsifier allows us to estimate this form only for $x \in \{0,1\}^{|V|}$. Most of the results in this paper only extend the notion of cut-sparsifiers, but not spectral sparsifiers. The only exception is for spectral sparsifiers of ``Cayley graphs'' on $\F_2^k$. In this special setting the vertex set of the Cayley graph is $\F_2^k$, and the edges are specified by a ``generating'' set $\Gamma \subseteq \F_2^k$. Two vertices $x, y \in \F_2^k$ are adjacent if $x-y \in \Gamma$. 

While the general theory of spectral sparsification of course holds for Cayley graphs, this may not lead to a compressed representation of the graph, since the generating set $\Gamma$ can be much smaller than $|V|$ (and $\Gamma$ specifies the Cayley graph completely). A natural question in this context would be whether there can be a compression of Cayley graphs that is also a Cayley graph (so leads to a compressed representation of the original graph). While this question remains open for general groups, in the setting of $\F_2^k$, our main theorem, \Cref{thm:main} effectively resolves this positively. 

A folklore connection between the eigenvectors of the Cayley graphs and the code generated by $\Gamma$ (where $\Gamma$ is viewed an $n \times k$ matrix over $\F_2$ whose columns generate a code contained in $\F_2^n$) allows us to show that the weight distribution of a code-sparsifier of $\Gamma$ closely matches that of $\Gamma$, and so leads to a new generating set for the Cayley graph with nearly matching eigenvalue profile. This leads to the following theorem, whose proof may be found in \Cref{sec:cayleyGraph}.
\begin{theorem}[Cayley Graph Spectral Sparsifier]\label{thm:cayleyIntro}
    For every (possibly weighted) Cayley graph $G$ on $\F_2^k$ with generating set $\Gamma \subseteq \F_2^k$, there exists a weighted sparsifier $\hat{\Gamma} \subseteq \Gamma$, such that for the Cayley graph $\hat{G}$ generated by $\hat{\Gamma}$,
    \[
    (1 - \eps) L_G \preceq L_{\hat{G}} \preceq (1 + \eps) L_G.
    \]
    Further, $|\hat{\Gamma}| \leq \widetilde{O}(k / \eps^2)$.
\end{theorem}

We stress that this is the first existential result of its kind. For comparison, the work of \cite{BSS09} showed the existence of spectral sparsifiers for any graph on $n$ vertices to size $O(n / \eps^2)$. In the setting of Cayley graphs, this implies the existence of spectral sparsifiers of Cayley graphs over $\F_2^k$ with $O(2^k / \eps^2)$ edges, leading to an average degree which is approximately $O(1 / \eps^2)$. \emph{However,} the key distinction is that the sparsifier returned by \cite{BSS09} is \emph{not} guaranteed to still be a Cayley graph and indeed it can not be. In fact, even just to maintain connectivity, a Cayley graph on $\F_2^k$ requires $\Omega(k)$ generators. Our sparsifiers have degree $\widetilde{O}(k)$ (for constant $\epsilon$), but now our resulting sparsified graph is a Cayley graph.

\paragraph{CSP Sparsification.}

As described earlier, cut sparsification can be viewed as a special case of CSP sparsification (corresponding to a CSP where constraints apply to two binary variables and require that their XOR be $1$). Furthermore when restricted to fields of constant size and generator matrices with exactly $r$ non-zero elements per row (where the columns of the generator matrix generate the code), the code sparsification problem is also a special case of CSP sparsification. Thus this interpretation already leads to a new broad class of CSPs that admit nearly linear sparsifiers. 

We say that a predicate $P:\F_q^r \to \{0,1\}$ is an {\em affine} predicate if there exist elements $a_0,a_1,\ldots,a_r \in \F_q$ such that $P(b_1,\ldots,b_r)=0$ if and only if $a_0 + \sum_i a_i b_i = 0$ (over $\F_q$). Equivalently, the predicate $P(b_1, \dots b_r)$ is evaluating $\sum_i a_i b_i \neq -a_0$. 

Now, let $\calP$ be a collection of predicates of any arity, and define CSP($\calP$) to be the family of CSPs where each constraint is a predicate from $\calP$ applied to any appropriately sized tuple of variables.

The following theorem asserts that CSPs over affine predicates are sparsifiable. 

\begin{theorem}\label{thm:affineCSPs}
    Let $\calP = \{P : P \text{ is an affine predicate over } \F_q \}$. Then, any CSP in CSP$(\calP)$ admits a nearly linear size $(1 \pm \eps)$ sparsification, namely of size $\widetilde{O}_q(k /\epsilon^2)$, where $k$ denotes the number of variables in the instance (and $O_q(\cdot)$ hides factors of $q$). 
\end{theorem}

Note that the above theorem has no dependence on the value $r$, and in fact, we can sparsify affine predicates even when $r = k$, and simultaneously sparsify affine predicates of different arities as long as they are affine with respect to the same field $\F_q$.

One immediate application of the above theorem is to any $r$-XOR constraint. Indeed, the unsatisfying instances of an XOR constraint form a linear subspace over $\F_2$, and so we give the first proof of the sparsifiability of XOR constraints to nearly linear size. This addresses one of the open questions of \cite{BZ20}, who showed the fundamental inexpressability of XOR constraints in terms of hypergraphs.

To illustrate the power of the theorem above, we also extend a result of \cite{FK17} to predicates on $3$ Boolean variables, giving an exact classification of which Boolean predicates on up to three variables are sparsifiable to near-linear size.  We say that a predicate $P: \zo^r \ra \zo$ has {\em an affine projection to AND} if there exists a function $\pi:[r] \to \{0,1,x,\neg{x}, y, \neg{y}\}$ such that $AND(x,y) = P(\pi(1),\ldots,\pi(r))$.    

\begin{theorem}\label{thm:ternaryCSPs}
	For a predicate $P: \zo^3 \ra \zo$, all possible CSPs of $P$ on subsets of $k$ variables are $(1 \pm \eps)$ sparsifiable to size $\widetilde{O}(k / \eps^2)$ if and only if $P$ has no affine projection to AND.
\end{theorem}

We remark that this classification does not yet extend to arbitrary ternary predicates (specifically over non-Boolean variables), and thus does not extend the result in \cite{BZ20}.

\subsection{Proof Techniques}

In our view, one of the strengths of this paper is that the proofs are conceptually simple and short even as they generalize known results and provide some new applications. Indeed once we determine the right form of the weight counting bound, namely \Cref{thm:newKargerIntro} (later as \Cref{cor:newKarger}), its proof is not very hard.

Roughly, we build a contraction procedure in linear spaces analogous to Karger's contraction method in \cite{Kar93, Kar99} in graphs. It is not immediately clear what should be the appropriate analog to Karger's contraction in our setting. Karger's method is inherently very reliant on the underlying graph structure, as each contraction \say{merges} two vertices with an edge between them. In a general linear code, while one can think of each row of the generating matrix as corresponding to an edge, and each column of the generating matrix as corresponding to a vertex, the analogy quickly breaks down as each row can be \say{involved} with many columns. As such, the perspective we end up taking for \say{contracting} the generating matrix (and one that extends to the graphical case for edges) is a contraction on a single coordinate $j$ of the generating matrix which is not always $0$. We decompose the entire column space of the generating matrix into the entire linear subspace of codewords ($\subseteq \F_q^n$) which is zero on this coordinate (i.e. such that the generating matrix for this subspace is entirely $0$ in the $j$th row), and a single vector which is non-zero in the $j$th coordinate. The original linear space is the span of these two separate components, and for us, contracting on a coordinate corresponds to keeping only the linear subspace which is zero on this coordinate.

Using this perspective, we show that this contraction procedure when applied to a random, not constantly $0$ coordinate of the code (which corresponds to an edge in the underlying graph in Karger's result) is likely to keep low-weight codewords intact. Thus if we focus on a particular low-weight codeword, repeated application of the contraction procedure should output this codeword with probability at least $q^{-\alpha} \binom{k}{\alpha}^{-1}$ \emph{as long as the support of our code never gets too small}. We use support here to refer to the total number of remaining not all zero rows of the generating matrix, or equivalently the number of coordinates in the code which are not always $0$.

While Karger's analysis is able to explicitly lower bound the support size (in his case the number of remaining edges) at every iteration in the algorithm by considering the minimum cut size, we can do no such thing as there can exist codes with large dimension and small support contained in our code. Instead, we build a parameterized trade-off saying that whenever our algorithm is unable to make progress in the contraction process, it is because there is a non-trivially high-dimensional code contained in our code that is supported on relatively few coordinates. (See \Cref{lem:newKarger} for a precise formulation.) Removing this code and the coordinates it is supported on, and continuing leads to a proof of \Cref{thm:newKargerIntro}.

The proof of \Cref{thm:main} given \Cref{thm:newKargerIntro} is also not hard, and we believe this is a simpler proof than previous sparsification results for special cases of graphs and hypergraphs. 
To prove the existence of near-linear size code sparsifiers, suppose we start with a code $\calC$ of dimension $k$, and length $k^2$. We first invoke the above theorem with $d = \sqrt{k}$. Intuitively, this breaks the code into two parts: In one part, we have no guarantee on the distribution of weights of the codewords, but the support size is bounded by $k \cdot d = k^{3/2}$. In the other part, we have a strong bound on the distribution of weights of codewords, but the support can still be as large as $k^2$. However, the fact that this code has roughly at most $k^\alpha$ codewords of weight $\alpha \sqrt{k}$ suggests that if we sample $k^{3/2}$ coordinates of this code and give each a weight of $\sqrt{k}$ then we get a pretty good sparsification of this part of the code. Gluing the two parts together gives an $O(k^{3/2})$ size sparsifier of the whole code. 
To get better sparsifications we can continue to apply this procedure recursively. For instance, if we repeat the process one more level (separately on each of the sparsified codes above), we can now use $d = k^{1/4}$ instead of $d = \sqrt{k}$ and this leads to $4$ codes, each of size roughly $k^{5/4}$. Likewise, we can glue these $4$ codes together, yielding an $O(k^{5/4})$ size sparsifier. Repeating enough times, and optimizing the parameters carefully leads to the final result. There are additional caveats when moving to codes whose length is super-polynomial in their dimension, and for these we utilize a few additional ideas to get our final sparsification result.

\subsection{Organization}

In \Cref{sec:newKargerBound}, we introduce an analog of Karger's contraction algorithm using Gaussian elimination to prove our decomposition theorem (Theorem~\ref{thm:newKargerIntro}) for linear codes. Then, in \Cref{sec:preliminaries} we introduce some basic facts and definitions about codes, graphs, and CSPs, which we will utilize in our construction of code sparsifiers and their applications. In \Cref{sec:polydimspars} we formalize the argument we gave above regarding sparsifying any code whose length is polynomial in its dimension. Finally, in \Cref{sec:anylengthsparse}, we use the algorithm from \Cref{sec:polydimspars} as a sub-routine along with several code decomposition tricks to remove any dependence on the length of the code, and present the existence of near-linear size sparsifiers for all linear codes, proving Theorem~\ref{thm:main}. In \Cref{sec:cayleyGraph}, we prove Theorem \ref{thm:cayleyIntro} by a simple reduction to codes. In \Cref{sec:csps} we prove Theorem \ref{thm:affineCSPs} and Theorem \ref{thm:ternaryCSPs}, extending the classification of near-linearly sparsifiable CSPs again by reducing to the coding case. Finally, in \Cref{sec:hypergraph}, we interpret hypergraphs in the setting of codes, proving our decomposition result, namely Theorem \ref{thm:hypergraphIntro}.

Finally, as a stand-alone application of our sparsification approach, we provide in \Cref{sec:simplerBK} of the appendix, a self-contained simpler proof of the near-linear size cut sparsifier result of~\cite{BK96}. 

\section{A Counting Bound for Codewords}\label[customsection]{sec:newKargerBound}

\subsection{Preliminaries}

First, we introduce a few basic definitions. These definitions will be used throughout the paper and are all that is required for this section.

\begin{Definition}
    A linear code $\calC$ of dimension $k$ and length $n$ is a $k$-dimensional linear subspace of $\F_q^n$. We often associate with $\calC$ a generator matrix $G \in \F_q^{n \times k}$, which maps vectors $x \in \F_q^k$ to codeword $\in \F_q^{n}$.
\end{Definition}

\begin{Definition}
    For a codeword $c \in \F_q^n$, the weight of a codeword is the number of non-zero entries in $c$. This is will be denoted as $\wt(c)$. 
\end{Definition}

\begin{Definition}
    For a linear code $\calC \subseteq \F_q^n$, we say the support size is 
    \[
    \Supp(\calC) = \left | \{ i \in [n]: \exists c \in \calC \text{ such that } c_i \neq 0 \} \right |. 
    \]
    In words, it is the number of coordinates of the code which are not always zero. Likewise, we say that for a code $\calC \subseteq \F_q^n$ along with a generating matrix $G \in \F_q^{n \times k}$, a coordinate $j \in [n]$ is \textbf{non-zero} if there exists a codeword $c \in \calC$ such that $c_j \neq 0$.
\end{Definition}

\begin{Definition}
    For a linear code $\calC \subseteq \F_q^n$, a \textbf{subcode} of $\calC$ is a linear subspace of $\calC$.
\end{Definition}

\begin{Definition}
    For a linear code $\calC \subseteq \F_q^n$, we say that the \textbf{density} of a code is \[
    \Density(\calC) = \frac{\text{Dim}(\calC)}{\Supp(\calC)}.
    \]
    We say that the density of the empty code is $0$.
\end{Definition}

\subsection{Overview}

In his seminal work, Karger~(\cite{Kar93, Kar99}) showed that for a graph on $k$ vertices with minimum cut-value $c$, there are at most $k^{2\alpha}$ cuts with size $\leq \alpha c$, for any integer $\alpha$. As discussed in the introduction, while an analogous statement does not hold for codes in general, we are able to prove a generalization that establishes a bound on the distribution of codewords. Roughly speaking, we show that for a code $\calC \subseteq \F_q^n$ of dimension $k$ and a parameter $d \in \mathbb{Z}^+$ of our choosing, either there exists a dense subcode of $\calC$ contained on a small support (i.e. a subcode whose density is $\geq \frac{1}{d}$), or the code satisfies a Karger-style bound on the distribution of codeword weights with parameter $d$ (i.e. there are at most $(qk)^{\alpha}$ codewords of weight $\leq \alpha d$). An exact statement of this result is given in Theorem \ref{lem:newKarger}. We will then show that Theorem \ref{lem:newKarger} in fact implies \Cref{thm:newKargerIntro} described in the introduction; that is, there exists a set of at most $kd$ coordinates, such that upon removing these coordinates (equivalently removing the corresponding rows from the generating matrix), the resulting code satisfies the Karger-style bound with parameter $d$. An exact statement is provided in Theorem \ref{cor:newKarger}.

 We will prove this bound by describing a \say{contraction} algorithm akin to that of Karger. Intuitively, the algorithm takes in a generating matrix of a code of dimension $k$ and length $n$. At random, the algorithm chooses one of the non-zero coordinates of this generating matrix and performs Gaussian elimination to \say{zero} out all but one of the columns in this coordinate. Then, the algorithm removes the remaining column of the generating matrix that is non-zero in this coordinate, reducing the dimension of the generating matrix by $1$. This process is repeated until the dimension of the generating matrix is sufficiently small. 
 
We call this Gaussian elimination step on a random non-zero coordinate of the generating matrix a \say{contraction}. We will show that as long as the support of the generating matrix is sufficiently large in every iteration, then any low-weight codeword will \say{survive} (i.e. remain in the span of the generating matrix) after many contractions with high probability. We can then conserve probability mass to argue that in fact, under the condition that the support is never too small, the number of lightweight codewords cannot be too large. This naturally leads to the statement of Theorem \ref{lem:newKarger} as either there exists a subcode of sufficiently high dimension with small support or during our contraction algorithm, the support is always large enough to get a strong bound on the number of codewords.

\begin{remark}
    When the mother code is the cut-code of a graph, the procedure of choosing a random row and eliminating all but $1$ of the non-zero entries is in fact equivalent to Karger's contraction based algorithm~(\cite{Kar93, Kar99}).
\end{remark}

First, we will prove some facts about the following algorithm, which takes as input a generating matrix $G \in \F_q^{n \times k}$ for a code of dimension $k$:

\begin{algorithm}\label{alg:Contraction}
    \caption{\text{Contract}$(G, \alpha)$}
    Let $G_i$ be the $i$th column of $G$, and let $k$ be the number of columns of $G$, and $n$ the number of rows.\\
    \While{$\text{dim}(G) \geq \alpha + 1$}{
    Choose a random non-zero coordinate $j \in [n]$ of $G$.\\
    Let $G_a$ be the first column of $G$ where the $j$th coordinate is non-zero, and let $G_{b_1}, \dots G_{b_p}$ be the remaining columns where the $j$th coordinate is non-zero. \\
    Remove column $G_a$ from $G$, and add $- G_{j,a}^{-1} G_{b_i, a} G_{a} $ to each $G_{b_i}, i \in [p]$.
    }
    \end{algorithm}

\begin{claim}
    Given a matrix $G$ of dimension $k$, after $i$ contractions, the dimension of the column span of $G$ is $k-i$.
\end{claim}

\begin{proof}
    For the base case, note that after $0$ contractions, the dimension is indeed $k$, as the columns of $G$ are indeed a basis. Now, suppose the claim holds inductively. We will assume that after $i$ contractions, the rank is $k-i$, and show that this holds after the $i+1$st contraction. This is clear however, as there are $k-i-1$ columns after the $i+1$st contraction, and if we added the column that we removed back into the matrix, the rank would still be $k-i$ (as every column operation was invertible). If adding one vector can bring the dimension to $k-i$, the dimension before adding the vector must be at least $k-i-1$.
\end{proof}

\begin{claim}
    For any non-zero coordinate $j \in [n]$ of $G$ that we contract to get $G'$, the span of $G'$ is exactly all codewords in the column span of $G$ that are zero in coordinate $j$.
\end{claim}

\begin{proof}
    First, note that if a code of dimension $k'$ is non-zero in some coordinate $j$, then it is non-zero in this coordinate in exactly $(q-1) \cdot q^{k'-1}$ codewords (and zero in this coordinate in exactly $q^{k'-1}$ codewords). After we contract on this $j$th coordinate, we get a new generating matrix $G'$ of dimension $k' -1$, where the $j$th row is all $0$. Finally, because $G'$ is made by adding columns of $G$ together, the span of $G'$ is contained in the span of $G$. This means that the span of $G'$ is a $k'-1$ dimensional subspace of $G$ where the $j$th row is $0$, which is exactly all codewords generated by $G$ that are $0$ in their $j$th coordinate, as the span of $G'$ will have $q^{k'-1}$ codewords.
\end{proof}

\begin{claim}\label{clm:stillInSpan}
    Consider a code $\calC$ of dimension $k$, and a codeword $c \in \calC$. If we never contract on any coordinate $j$ where $c$ is non-zero, then $c$ is still in the span of the resulting contracted code. 
\end{claim}

\begin{proof}
    Let $G'$ be the result of performing all the aforementioned contractions. Let the sequence of coordinates we contract on be ${j_1}, \dots {j_k}$. We know that after each contraction on ${j_i}$, the span of the new generator matrix is exactly all remaining codewords of $G$ that are zero in coordinate ${j_i}$. Since $c$ is zero on all of ${j_1}, \dots {j_k}$ (because we only ever contract on coordinates where $c$ is 0 by assumption), then $c$ always remains in the span of $G$ after each contraction as the columns we removed are non-zero on these coordinates. So, after all the contractions are performed, $c$ is still in the span of $G$. 
\end{proof}

\subsection{A Karger-Style Bound for Codewords}

Here, we will prove the following theorem:

\begin{theorem}\label{lem:newKarger}
    For a linear code $\calC \subseteq \F_q^n$ of dimension $k$, and any integer $d \geq 1$, at least one of the following is true:
    \begin{enumerate}
        \item There exists a linear sub-code $\calC' \subset \calC$ such that $\Density(\calC') > \frac{1}{d}$.
        \item For all integers $\alpha$, there are at most $q^{\alpha} \cdot \binom{k}{\alpha}$ codewords of weight $\leq \alpha d$.
    \end{enumerate}
\end{theorem}

\begin{remark}
    Note that Theorem \ref{lem:newKarger} implies Karger's original cut-counting bound as a special case with $q=2$. For any $c \geq 1$, in any graph with minimum cut size $c$, the number of edges is necessarily at least $(nc)/2$. So Condition $1$ above never arises once we set $d = c/2$, allowing us to recover Karger's original cut-counting bound.
\end{remark}

We first prove some sub-claims that will make this easier.

\begin{lemma}\label{clm:probSurvive}
    Suppose we run Algorithm \ref{alg:Contraction}, and for some $d \in \mathbb{Z}^+$, after every contraction $\Density(\Span(G)) \leq \frac{1}{d}$. Then, the probability some codeword $c \in \calC$ of weight $\leq \alpha d$ (for $\alpha \in \mathbb{Z}^+)$ is still in the span of the final contracted $G$ is at least $\binom{k}{\alpha}^{-1}$.
\end{lemma}

\begin{proof}
     By Claim \ref{clm:stillInSpan}, it follows that if $c$ is in the span of the generating matrix $G$ after $i$ iterations, and we contract on a coordinate where $c$ is non-zero, then $c$ will still be in the span of the contracted generating matrix. So, it follows that the probability $c$ survives is:
     \[
     \Pr[\text{survives first contraction}] \cdot \dots \cdot \Pr[\text{survives } (k - \alpha) \text{th contraction}].
     \]
     Using the fact that before the $i$th contraction, the dimension is $k-i+1$, we know that the support must be at least $(k-i+1) \cdot d$ by the assumed density. Because the codeword $c$ survives if we contract on a coordinate where $c$ is $0$, the probability of survival in the $i$th contraction is at least $1 - \frac{\alpha \cdot d}{(k-i+1) \cdot d}$. Thus, 
     \begin{align}
         \Pr[\text{survives all contractions}] &\geq (1 - \alpha / k) (1 - \alpha / (k-1)) \dots (1 - \alpha / (\alpha+1)) \\
         &= \frac{k - \alpha}{k} \cdot \frac{k - 1 - \alpha}{k-1} \cdot \dots \cdot \frac{\alpha + 1 - \alpha}{\alpha+1} = \binom{k}{\alpha}^{-1}.
     \end{align}
\end{proof}

We are now ready to prove our main claim. 

\begin{proof}[Proof of Theorem \ref{lem:newKarger}]
    Suppose that condition 1 does not hold. Then, every linear sub-code $\calC' \subseteq \calC$ satisfies $\Density(\calC') \leq \frac{1}{d}$. We can then invoke Lemma \ref{clm:probSurvive} to conclude that any codeword with weight $\leq \alpha \cdot d$ is in the span of the contracted matrix with probability $\geq \binom{k}{\alpha}^{-1}$. Further, note that because the dimension of the contracted matrix is $\alpha$, there are at most $q^{\alpha}$ codewords in this span. Because there are $q^{\alpha}$ codewords in the span of each contracted matrix, the sum of all the probabilities of low weight codewords surviving must be $\leq q^{\alpha}$. This means there can be at most $q^{\alpha} \cdot \binom{k}{\alpha}$ codewords of weight $\leq \alpha \cdot d$. 
\end{proof}

\begin{theorem}\label{cor:newKarger}
    For a linear code $\calC$ of dimension $k$ and length $n$ over $\F_q$, for any integer $d \geq 1$, there exists a set of at most $k \cdot d$ coordinates, such that upon their removal, in the resulting code, for any integer $\alpha \geq 1$ there are at most $q^{\alpha} \cdot \binom{k}{\alpha}$ codewords of weight $\leq \alpha d$.
\end{theorem}

\begin{proof}
    As long as Condition 1 of Theorem \ref{lem:newKarger} continues to hold, we can write the matrix in the form $\begin{bmatrix}
        A & B \\
        0 & C 
    \end{bmatrix}$, where the coordinates corresponding to $A, B$ are the subcode of density $> \frac{1}{d}$. We can then remove all these coordinates corresponding to this subspace, and continue repeating this process until Condition 1 no longer holds. Once condition 2 holds, the dimension of the new code will be some value $\leq k$, so the number of codewords of weight $\leq \alpha d$ will be at most $q^{\alpha} \cdot \binom{k}{\alpha}$. To see why the number of coordinates we remove is at most $kd$, whenever the subspace we remove has dimension $k'$, we remove at most $k'd$ coordinates, so after this removal, the resulting code is of dimension $k-k'$. It follows that we can remove at most $kd$ coordinates before the dimension of the code is $0$.
    
    Note that after removing coordinates, the resulting code will have dimension $\leq k$, so in particular, the generating matrix will have a non-trivial nullspace. This means that there will be several messages that map to the \emph{same} codeword. However, if the encoding of two messages is the same, i.e. yielding the same codeword, we do not count these as separate instances. Instead, this bound treats this as a single codeword.
\end{proof}

\section{Preliminaries}\label[customsection]{sec:preliminaries}

\subsection{Codes}

\begin{Definition}
    For a code $\calC \subseteq \F_q^n$, its \textbf{distance} is 
    \[
    \min_{c \in \calC: c \neq 0} \wt(c).
    \]
\end{Definition}

\begin{Definition}
    For a code $\calC \subseteq \F_q^n$, the \textbf{coordinates} of the code are $[n]$. When we refer to the \emph{number of coordinates}, this is interchangeable with the \emph{length} of the code, which is exactly $n$.
\end{Definition}

In this work, we will be concerned with code sparsifiers as defined below.

\begin{Definition}
    For a code $\calC \subseteq \F_q^n$ with associated generating matrix $G$, a $(1 \pm \eps)$-sparsifier for $\calC$ is a subset $S \subseteq n$, along with a set of weights $w_S: S \ra \mathbb{R}^{+}$ such that for any $x \in \F_q^k$
    \[
    (1 - \eps) \wt(Gx) \leq \wt_S(G|_S x) \leq (1 + \eps) \wt(Gx).
    \]

    Here, $\wt_S$ is meant to imply that if the codeword is non-zero in its coordinate corresponding to an element $i \in S$, then it contributes $w_S(i)$ to the weight. We will often denote $G|_S$ with the corresponding weights as $\widetilde{G}$.
\end{Definition}

We next present a few simple results for code sparsification that we will use frequently.

\begin{claim}\label{clm:verticalDecomp}
    For a vertical decomposition of a generating matrix, \[
    G = \begin{bmatrix}
        G_1 \\
        G_2 \\
        \vdots \\
        G_k
    \end{bmatrix},
    \]
    if we have a $(1 \pm \eps)$ sparsifier to codeword weights in each $G_i$, then their union is a $(1 \pm \eps)$ sparsifier for $G$.
\end{claim}
\begin{proof}
    Consider any codeword $c \in \text{Span}(G)$. Let $c_i$ denote the restriction to each $G_i$ in the vertical decomposition. It follows that if in the sparsifier $\wt(\hat{c}_i) \in (1 \pm \eps) \wt(c_i)$, then $\wt(\hat{c}) = \sum_i\wt(\hat{c}_i) \in (1 \pm \eps) \sum_i \wt(c_i)  = (1 \pm \eps) \wt(c)$. 
\end{proof}

\begin{claim}\label{clm:composingApproximations}
Suppose $\calC'$ is $(1 \pm \delta)$ sparsifier of $\calC$, and $\calC''$ is a $(1 \pm \eps)$ sparsifier of $\calC'$, then $\calC''$ is a $(1 - \eps)(1-\delta), (1 + \eps)(1 + \delta)$ approximation to $\calC$ (i.e. preserves the weight of any codeword to a factor $(1 - \eps)(1 - \delta)$ below and $(1 + \eps)(1 + \delta)$ above).
\end{claim}

\begin{proof}
    Consider any codeword $\calC x$. We know that $(1 - \eps)\wt(\calC x) \leq \wt(\calC'x) \leq (1 + \eps)\wt(\calC x)$. Additionally, $(1 - \delta)\wt(\calC'x) \leq \wt(\calC''x) \leq (1 + \delta)\wt(\calC'x)$. Composing these two facts, we get our claim. 
\end{proof}

Finally, we will use the following claim many times implicitly in our arguments, as we will freely change the generating matrix of the code we are looking at. 

\begin{claim}
    Suppose generating matrices $G$ and $G'$ both generate the same dimension $k$ code $\calC$ of length $n$. Then, if some weighted subset of the rows of $G$ yields a $(1 \pm \eps)$ sparsifier $G|_S = \hat{G}$, the same weighted subset of the rows of $G'$ yields a $(1 \pm \eps)$ sparsifier $G'|_S = \hat{G'}$.
\end{claim}

\begin{proof}
    Consider any codeword $c \in \calC$. By construction, there is an $x, x'$ such that $Gx = c, G'x' = c$. Now, $G|_S x = c|_S$ and $G'|_S x' = c|_S$. Hence, if $\hat{G}$ is a $(1 \pm \eps)$ sparsifier of codewords in $\calC$, then so too is $\hat{G'}$.
\end{proof}

\subsection{A Probabilistic Bound}

We will frequently utilize the following probabilistic bound. 

\begin{claim}\label{clm:concentrationBound}{\rm (\cite{FHH11})}
    Let $X_1, \dots X_{\ell}$ be random variables such that $X_i$ takes on value $1 / p_i$ with probability $p_i$, and is $0$ otherwise. Also, suppose that $\min_i p_i \geq p$. Then, with probability at least $1 - 2e^{-0.38 \eps^2 \ell p}$,
    \[
    \sum_i X_i \in (1 \pm \eps) \ell.
    \]
\end{claim}

\subsection{Graphs and Graph Sparsification}\label[customsection]{sec:graphs}

We recall here a few basic concepts about graphs and graph sparsification. 

\begin{Definition}
    A \emph{cut} in a graph $G = (V, E)$ is a subset $S \subseteq V$. We will specify the size of a cut $|\delta_G(S)|$ to be the number of edges that cross from $S$ to $V - S$ (i.e. the number of edges that go from a set $S$ to the rest of the vertices). When a graph $G = (V, E)$ also has an associated weight function $w: E \ra \mathbb{R^+}$, we take $|\delta_G(S)|$ to be the sum over all the edges that cross from $S$ to $V-S$ of the weights of these crossing edges.
\end{Definition}

\begin{Definition}
    A $(1 \pm \eps)$ \emph{cut}-sparsifier for a weighted graph $G = (V, E)$ is a new, weighted graph $\hat{G} = (V, \hat{E})$, with associated weight function $w$, such that
    \begin{enumerate}
        \item $\hat{E} \subseteq E$.
        \item For every cut $S \subseteq V$, 
        \[
        (1 - \eps) |\delta_G(S)| \leq |\delta_{\hat{G}}(S)| \leq (1 + \eps) |\delta_G(S)|.
        \]
    \end{enumerate}
\end{Definition}

The famous result of Karger relates the size of the minimum cut to the number of cuts of other sizes.

\begin{Definition}
    The \emph{minimum cut} of a graph $G = (V, E)$ is
    \[
    \min_{S \subseteq V: S \neq V, S \neq \emptyset} |\delta_G(S)|.
    \]
\end{Definition}

\begin{theorem}[Karger's Cut-Counting Bound]\cite{Kar93, Kar99}\label{thm:kargerCutCounting}
    Suppose a graph $G = (V, E)$ has $n$ vertices, and minimum cut value $c$. Then, for any integer $\alpha$, the number of cuts of size at most $\alpha c$ is at most $n^{2 \alpha}$.
\end{theorem}

\begin{Definition}[Cut Code]
For intuition, we will several times refer to the \say{cut code} of a corresponding graph $G = (V, E)$. Intuitively, this is the code over $\F_2$ with a generating matrix on $|V|$ columns, where for every edge $e = (u, v) \in E$, we add a row in the generating matrix which has a $1$ in the columns corresponding to $u$ and $v$. If we denote this generating matrix by $G'$, it can be verified that for any cut $(S, V-S)$, 
\[
 | \delta_G(S) | = \wt(G' \mathbf{1}_S).
\]
That is, the weight of the codeword corresponding to the encoding of the indicator vector of $S$ is exactly the number of edges crossing the cut $(S, V-S)$. 
\end{Definition}

\subsection{CSPs and CSP Sparsification}

We formally define here CSPs and CSP sparsification. We start by introducing the notion of a predicate.

\begin{Definition}
	A predicate $P$ of arity $r$ is a function going from $\zo^r \ra \zo$. 
\end{Definition}

We will look at CSPs which are defined using a single predicate.

\begin{Definition}
	A CSP over $k$ variables with predicate $P$, is a collection of constraints of the form $P(x^{(i)}_{1}, x^{(i)}_{2}, \dots x^{(i)}_{r} )$ where $r$ is the arity of $P$, and $i$ ranges from $1$ to $m$. 
\end{Definition}

\begin{Definition}
	The value obtained by the CSP on an assignment $x$ is correspondingly
	\[
	\sum_{i = 1}^m P(x^{(i)}_{1}, x^{(i)}_{2}, \dots x^{(i)}_{r} )
	\]
 In some cases, the CSP has a corresponding \emph{weight} $w_i \in \mathbb{R}^+$ for each constraint. In this case, the value of the CSP on assignment $x$ is
 	\[
	\sum_{i = 1}^m w_i \cdot P(x^{(i)}_{1}, x^{(i)}_{2}, \dots x^{(i)}_{r} ).
	\]
\end{Definition}

\begin{Definition}
	A $(1 \pm \eps)$-sparsifier for a CSP $C$ on $m$ constraints, is a new CSP $\hat{C}$ specified by a subset $T \subseteq [m]$, along with weights $(w_i)_{i \in T}$, such that for any assignment $x \in \zo^k$,
	\[
	(1 - \eps) \sum_{i = 1}^m P(x^{(i)}_{1}, x^{(i)}_{2}, \dots x^{(i)}_{r} )  \leq \sum_{i \in T} w_i P(x^{(i)}_{1}, x^{(i)}_{2}, \dots x^{(i)}_{r} ) \leq (1 + \eps) \sum_{i = 1}^m P(x^{(i)}_{1}, x^{(i)}_{2}, \dots x^{(i)}_{r} ) .
	\]
	In words, we choose a weighted subset of the constraints of $C$, such that for any assignment $x$, the value of the CSP is preserved to a $(1 \pm \eps)$ factor. 
\end{Definition}

Note that in some works, sparsifying a CSP is meant to only preserve satisfiability while reducing the number of constraints. In our setting, the goal is to approximately preserve the value of the satisfied constraints.

\begin{Definition}\label{def:affineProjection}
	For a universe of variables $x \in \zo^k$, an affine projection is a restriction of the variables of the form $x_i = 0, x_i = 1, x_i = x_j$ or $x_i = \neg x_j$. 
\end{Definition}

\begin{remark}
	We say that an affine projection of a predicate $P$ of arity $r$ yields an AND of arity $2$ if there exists a function $\pi:[r] \to \{0,1,x,\neg{x}, y, \neg{y}\}$ such that $\mathrm{AND}(x,y) = P(\pi(1),\ldots,\pi(r))$.
	
	For instance, the predicate $P: \zo^3 \ra \zo$, with the only satisfying assignments $000, 001$ is equal to an AND of arity $2$ under affine projections. If we consider the restriction $R$ which sets the third variable equal to $0$, we get a new predicate $P|_R$ whose only satisfying assignment is $00$. Thus, this predicate $P|_R(y_1, y_2)  = \neg y_1 \wedge \neg y_2$.
\end{remark}

\section{Near-linear Size Sparsifiers for Polynomially-bounded Codes}\label[customsection]{sec:polydimspars}

In this section, we will prove the existence of near-linear size sparsifiers for codes of length $k^{O(1)}$, where $k$ is the dimension of the code. That is, when $\calC$ is a linear code of dimension $k$ and length $k^{O(1)}$.  We will also assume that $\calC$ is unweighted (or that every coordinate has the same weight). The main theorem to be proved in this section is Theorem \ref{thm:sparsifyCodesPolyLength} showing the result for polynomially length (which is a specific case of the more general Theorem \ref{thm:codeSparsifyGeneralLength} which applies to codes of any length, but loses extra factors, also proved here). In particular, this is a special case of \Cref{thm:main} proved in the introduction.

Intuitively, the proof will use Theorem \ref{cor:newKarger} to repeatedly decompose the code $\calC$. In each application, we invoke Theorem \ref{cor:newKarger} with a specific choice of parameter $d$ to decompose the generator matrix for the code $\calC$ into the form
\[
\begin{bmatrix}
    A & B \\
    0 & C 
\end{bmatrix},
\]
where $A$ is exactly the low-dimensional subcode with bounded support. We will argue that we can keep all of the coordinates of $A$ because the support of $A$ is sufficiently bounded. In turn, this means that $B$ is effectively preserved as well, so as long as we can get a $(1 \pm \eps)$ approximation to $C$ of sufficiently small size, we will be okay. To deal with $C$, we argue that the distribution of the weights of codewords in $C$ is sufficiently smooth, such that we may subsample the coordinates at rate roughly $1/d$. 

This now yields two separate codes, each with size that is strictly smaller than the starting size. We operate inductively, and recursively break down these two smaller codes. Ultimately, in each recursive step, we take a code of size $k \cdot k^{\gamma}$, and return two codes of size roughly $k \cdot k^{\gamma / 2}$. For an initial code of length $k^{O(1)}$ (i.e. whose length is polynomial in the dimension), after $\log \log k$ levels of recursion, we have $\log k$ codes, each of length roughly $k$. In turn, we can glue these codes back together, and return a sparsifier for our original code.

\subsection{Code Decomposition}

First, we consider Algorithm \ref{alg:CodeDecomposition}:

\begin{algorithm}\label{alg:CodeDecomposition}
\caption{CodeDecomposition$(\calC, d)$}
Let $k$ be the dimension of $\calC$. \\
Let $S$ be the set of coordinates to be removed as specified by Theorem \ref{cor:newKarger}. \\
Let $\calC'$ be the code $\calC$ after removing the set of coordinates $S$. \\
\Return{$S, \calC'$}
\end{algorithm}

\begin{claim}
\begin{enumerate}
    \item After the termination of the Algorithm \ref{alg:CodeDecomposition}, $|S| \leq k \cdot d$.
    \item The final resulting $\calC'$ of Algorithm \ref{alg:CodeDecomposition} satisfies condition 2 of Theorem \ref{lem:newKarger}.
\end{enumerate}
    
\end{claim}

\begin{proof}
For item 1:
    This follows exactly from Theorem \ref{cor:newKarger}.

For item 2:
    Because we exited the while loop, Condition 1 no longer holds for $\calC'$. Thus, by Theorem \ref{lem:newKarger}, Condition 2 must hold.
\end{proof}

\begin{claim}
    To get a $(1\pm \eps)$ code sparsifier for $\calC$, it suffices to get $S, \calC'$ from Algorithm \ref{alg:CodeDecomposition}, and then sample all of the indices in $S$ with probability $1$, and get a $(1 \pm \eps)$-sparsifier for $\calC'$.
\end{claim}

\begin{proof}
    This follows because we are creating a \say{vertical} decomposition of the code. The coordinates of the code correpsonding to $S$ contain a dimension $k'$ subspace, and the remaining coordinates of the code define $\calC'$. Thus, we conclude by using Claim \ref{clm:verticalDecomp}.
\end{proof}

\subsection{Code Sparsification Algorithm}

In this section we will present the code sparsification algorithm, and argue its correctness and its sparsity. 

\begin{algorithm}
    \caption{CodeSparsify$(\calC \subseteq \F_q^{n}, k, \eps, \eta)$}\label{alg:CodeSparsify}
    Let $n$ be the length of $\calC$. \\
    \If{$n \leq 100 \cdot k \cdot \eta \log (k)\log(q) / \eps^2$}{\Return{$\calC$}}
    Let $d = \frac{n \eps^2}{\eta \cdot k \log (k) \log(q)}$. \\
    Let $S, \calC' = \text{CodeDecomposition}(\calC, \sqrt{d} \cdot \eta \cdot \log(k) \log(q) / \eps^2)$.
    Let $\calC_1 = \calC|_S$.
    Let $\calC_2$ be the result of sampling every coordinate of $\calC'$ at rate $1 / \sqrt{d}$. \\
    \Return{$\mathrm{CodeSparsify}(\calC_1, k, \eps, \eta) \cup \sqrt{d} \cdot \mathrm{CodeSparsify}(\calC_2, k, \eps, \eta)$ }
\end{algorithm}

We will use the following fact, which is a simple extension of a result from Karger \cite{Kar94sparsification}. In Karger's work, it was noted that for a graph with minimum cut value $c$, one can roughly sample the edges at rate $\log(n) / (c \eps^2)$ and scale the weights of the sampled edges up by $c\eps^2 / \log(n)$ while still maintaining a $(1 \pm \eps)$ approximation to the cuts in the graph. In the following claim, we adapt this fact to codes which satisfy a smooth bound on the number of codewords of a given weight.

\begin{claim}\label{clm:preserveWeightSubsample}
    Suppose $\calC$ is a code of dimension $k$ over $\F_q$, and let $b \geq 1$ be an integer such that for any integer $\alpha \geq 1$, the number of codewords of weight $\leq \alpha b$ is at most $(qk)^{\alpha}$. Suppose further that the minimum distance of the code $\calC$ is $b$. Then, sampling the coordinates of $\calC$ at rate $\frac{\log(k) \log(q) \eta}{b \eps^2}$ with weights $\frac{b \eps^2}{\log(k) \log(q) \eta}$ yields a $(1 \pm \eps)$ sparsifier with probability $1 - 2^{-(0.19\eta - 110) \log k} \cdot k^{-101}$.
\end{claim}

\begin{proof}

    Consider any codeword $c$ of weight $[\alpha b / 2, \alpha b]$ in $\calC$. We know that there are at most $(qk)^{\alpha}$ codewords that have weight in this range. The probability that our sampling procedure fails to preserve the weight of $c$ up to a $(1 \pm \eps)$ fraction can be bounded by Claim \ref{clm:concentrationBound}. Indeed,
    \[
    \Pr[\text{fail to preserve weight of } c] \leq 2e^{-0.38 \cdot \eps^2 \cdot \frac{\alpha b}{2} \cdot \frac{\eta \log (k) \log(q)}{\eps^2 b}} = 2e^{-0.19 \alpha \eta \log (k)\log(q)}.
    \]
    Now, let us take a union bound over the at most $(qk)^{\alpha}$ codewords of weight between $[\alpha d / 2, \alpha]$. Indeed,
    \begin{align*}
    \Pr[\text{fail to preserve any } c \text{ of weight } [\alpha b / 2, \alpha b]] &\leq 2^{\alpha \log (qk)} \cdot 2e^{-0.19 \alpha \eta \log (k)\log(q)} \\
     & \leq 2^{\alpha \cdot (-0.19\eta + 1) \log (k)\log(q)} \\
     & \leq 2^{\alpha \cdot (-0.19\eta + 1) \log (k)} \\
     & \leq 2^{-(0.19\eta - 110) \alpha \log k} \cdot 2^{-109 \alpha \log k} \\
     & \leq 2^{-(0.19\eta - 110) \log k} \cdot k^{-109 \alpha},
    \end{align*}
    where we have chosen $\eta$ to be sufficiently large. Now, by integrating over $\alpha \geq 1$, we can bound the failure probability for any integer choice of $\alpha$ by $2^{-(0.19\eta - 110) \log k} \cdot k^{-101}$.
\end{proof}

\begin{lemma}\label{clm:overallSpace}
    In Algorithm \ref{alg:CodeSparsify}, starting with a code $\calC$ of size $d k \log (k) \log(q) / \eps^2$, after $i$ levels of recursion, with probability $1 - 2^{i} \cdot 2^{-\eta k}$, the code being sparsified at level $i$, $\calC^{(i)}$ has at most 
    \[
    (1 + 1 / 2 \log \log (k))^{i} \cdot d^{1/2^i} \cdot \eta \cdot k \log(k)\log(q) / \eps^2
    \]
    surviving coordinates.
\end{lemma}

\begin{proof}
    Let us prove the claim inductively. For the base case, note that in the $0$th level of recursion the number of surviving coordinates in $\calC^{(0)} = \calC$ is $d \cdot k \log (k) \log(q) / \eps^2$, so the claim is satisfied trivially.

    Now, suppose the claim holds inductively. Let $\calC^{(i)}$ denote a code that we encounter in the $i$th level of recursion, and suppose that it has at most \[
    (1 + 1 / 2 \log \log (k))^{i} \cdot d^{1/2^i} \cdot \eta \cdot k \log(k) \log(q) / \eps^2
    \]
    coordinates. Denote this number of coordinates by $\ell$. Now, if this number is smaller than $100 k \eta \log (k) \log(q) / \eps^2$, we will simply return this code, and there will be no more levels of recursion, so our claim holds vacuously. Instead, suppose that this number is larger than $100 k \eta \log (k) \log(q) / \eps^2$. Let $d' = \frac{\ell \eps^2}{\eta k \log(k) \log(q)} \leq (1 + 1 / 2 \log \log (k))^{i} \cdot d^{1/2^i}$.

    Then, we decompose $\calC^{(i)}$ into two codes, $\calC_1$ and $\calC_2$. $\calC_1$ contains coordinates of $\calC^{(i)}$ that contain some $k'$ dimensional code on a support of size $\leq k' \cdot \sqrt{d'} \cdot \eta \cdot \log(k) \log(q) / \eps^2$. Because $k' \leq k$, it follows that the number of coordinates in $\calC_1$, the first code we recurse on, is at most $(1 + 1 / 2 \log \log (k))^{i} \cdot d^{1/2^{i+1}} \cdot \eta \cdot k \log(k) \log(q) / \eps^2$ as we desire.

    For $\calC_2$, we define random variables $X_1 \dots X_{\ell}$ for each coordinate in the support of $\calC_2$. $X_i$ will take value $1$ if we sample coordinate $i$, and it will take $0$ otherwise. Let $X= \sum_{i = 1}^{\ell} X_i$, and let $\mu = \E[X]$. Note that 
    \[
    \frac{\mu^2}{\ell} = \left ( \frac{\ell}{\sqrt{d'}} \right )^2 / {\ell} = \frac{\ell}{d'} \geq \eta \cdot k \cdot \log(k) \log(q) / \eps^2.
    \]

    Now, using Chernoff, \[
    \Pr[X \geq (1 + 1 / 2 \log \log (k)) \mu] \leq e^{\frac{-2}{4 \log^2 \log(k)} \cdot \eta \cdot k \cdot \log(k) \log(q) / \eps^2} \leq 2^{-\eta k}, 
    \]
    as we desire. Since $\mu = \ell / \sqrt{d'} \leq (1 + 1 / 2 \log \log (k))^{i} \cdot d^{1/2^{i+1}} \cdot \eta \cdot k \log(k) \log(q) / \eps^2$, we conclude our result. 
    
    Now, to get our probability bound, we also operate inductively. Suppose that up to recursive level $i-1$, all sub-codes have been successfully sparsified to their desired size. At the $i$th level of recursion, there are at most $2^{i-1}$ codes which are being probabilistically sparsified. Each of these does not exceed its expected size by more than the prescribed amount with probability at most $2^{-\eta k}$. Hence, the probability all codes will be successfully sparsified up to and including the $i$th level of recursion is at least $1 - 2^{i-1} 2^{-\eta k} - 2^{i-1}2^{-\eta k} = 1 - 2^i 2^{-\eta k}$. 
\end{proof}

\begin{lemma}\label{clm:goodApprox1Iter}
For any iteration of Algorithm \ref{alg:CodeSparsify} called on a code $\calC$, $\calC_1 \cup \sqrt{d} \cdot \calC_2$ is a $(1 \pm \eps)$ approximation to $\calC$ with probability at least $1 - 2^{-(0.19\eta - 110) \log k} \cdot k^{-101}$.
\end{lemma}

\begin{proof}
    First, we note that $\calC', \calC|_S$ (as returned from Algorithm \ref{alg:CodeDecomposition}) form a \emph{vertical} decomposition of $\calC$. Hence, it suffices to show that $\calC_1$ is a $(1 \pm \eps)$-sparsifier to $\calC|_S$, and $\calC_2$ is a $(1 \pm \eps)$-sparsifier to $\calC'$. 

    Seeing that $\calC_1$ is a $(1 \pm \eps)$-sparsifier to $\calC|_S$ is trivial, as $\calC_1 = \calC|_S$, since we preserve every coordinate with probability $1$.

    To see that $\calC_2$ is a $(1 \pm \eps)$-sparsifier to $\calC'$, first note that every codeword in $\calC'$ is of weight at least $\sqrt{d} \cdot \eta \cdot \log(k) \log(q) / \eps^2$. This is because if there were a codeword of weight smaller than this, there would exist a subcode of $\calC'$ with dimension $1$, and support bounded by $\sqrt{d} \cdot \eta \cdot \log(k) \log(q) / \eps^2$. But, because we used Algorithm \ref{alg:CodeDecomposition}, we know that there can be no such sub-code remaining in $\calC'$. Thus, every codeword in $\calC'$ is of weight at least $\sqrt{d} \cdot \eta \cdot \log(k) \log(q) / \eps^2$. 

    Now, we can invoke Claim \ref{clm:preserveWeightSubsample} with $b = \sqrt{d} \eta \log(k) \log(q) / \eps^2$. Note that the hypothesis of Claim \ref{clm:preserveWeightSubsample} is satisfied by virtue of our code decomposition. Indeed, we removed coordinates of the code such that in the resulting $\calC_2$, for any $\alpha \geq 1$, there are at most $(qk)^{\alpha}$ codewords of weight $\leq \alpha \sqrt{d} \eta \log(k) \log(q) / \eps^2$. Using the concentration bound of Claim \ref{clm:preserveWeightSubsample} yields that with probability at least $1 - 2^{-(0.19\eta - 110) \log k} \cdot k^{-101}$, the resulting sparsifier for $\calC_2$ is a $(1 \pm \eps)$ sparsifier, as we desire.
    
\end{proof}

\begin{corollary}\label{clm:overallCodeAccuracy}
    If Algorithm \ref{alg:CodeSparsify} achieves maximum recursion depth $\ell$ when called on a matrix $\calC$, and $\eta > 600$, then the result of the algorithm is a $(1 \pm \eps)^{\ell}$ sparsifier to $\calC$ with probability $\geq 1 - (2^{\ell}-1) \cdot2^{-(0.19\eta - 110) \log k} \cdot k^{-101}$
\end{corollary}

\begin{proof}
    We prove the claim inductively. Clearly, if the maximum recursion depth reached by the algorithm is $0$, then we have simply returned the code itself. This is by definition a $(1 \pm \eps)^{0}$ sparsifier to itself.

    Now, suppose the claim holds for maximum recursion depth $i - 1$. We will show it holds for maximum recursion depth $i$. Let the code we are sparsifying be $\calC$. We break this into $\calC_1$, $\calC_2$, and sparsify these. By our inductive claim, with probability $1 - (2^{i-1}-1) \cdot 2^{-(0.19\eta - 110) \log k} \cdot k^{-101}$ each of the sparsifiers for $\calC_1, \calC_2$ are $(1 \pm \eps)^{i-1}$ sparsifiers. Now, by Lemma \ref{clm:goodApprox1Iter} and our value of $\eta$, $\calC_1, \calC_2$ themselves together form a $(1 \pm \eps)$ sparsifier for $\calC$ with probability $1 - 2^{-(0.19\eta - 110) \log k}\cdot k^{-101}$. So, by using Claim \ref{clm:composingApproximations}, we can conclude that with probability $1 - (2^i -1) \cdot 2^{-(0.19\eta - 110) \log k} \cdot k^{-101}$, the result of sparsifying $\calC_1, \calC_2$ forms a $(1 \pm \eps)^i$ approximation to $\calC$, as we desire.
\end{proof}

We can then state the main theorem from this section:

\begin{theorem}\label{thm:sparsifyCodesPolyLength}
    For a code $\calC$ on alphabet $\F_q$ of dimension $k$, and length $k^{O(1)}$, Algorithm \ref{alg:CodeSparsify} creates a $(1 \pm \eps)$ sparsifier for $\calC$ of size $O(k \eta \log^2(k) \log(q) (\log \log (k))^2 / \eps^2)$ with probability $1 - 2^{-(0.19\eta - 110) \log k} \cdot k^{-100}$.
\end{theorem}

\begin{proof}
    For a code of dimension $k$, and length $k^{O(1)}$, this means that our value of $d$ as specified in the first call to Algorithm \ref{alg:CodeSparsify} is at most $k^{O(1)}$ as well. As a result, after only $\log \log k$ iterations, $d = k^{O(1) / 2^{\log \log k}} = k^{O(1) / \log k} = O(1)$. So, by Corollary \ref{clm:overallCodeAccuracy}, because the maximum recursion depth is only $\log \log k$, it follows that with probability at least $1 - (2^{\log \log k} -1) \cdot 2^{-(0.19\eta - 110) \log k} \cdot k^{-101} \geq 1 - (\log k) \cdot 2^{-(0.19\eta - 110) \log k} \cdot k^{-101}$, the returned result from Algorithm \ref{alg:CodeSparsify} is a $(1 \pm \eps)^{\log \log k}$ sparsifier for $\calC$.

    Now, by Lemma $\ref{clm:overallSpace}$, with probability $\geq 1 - 2^{\log \log k} \cdot 2^{-\eta k} \geq 1 - \log (k) 2^{-\eta k} \geq 1 - 2^{-(0.19\eta - 110) \log k} \cdot k^{-101}$, every code at recursive depth $\log \log k$ has at most 
    \[
    (1 + 1 / 2 \log \log (k))^{\log\log k} \cdot d^{1/\log k} \cdot \eta \cdot k \log(k) \log(q) / \eps^2 = O(k \eta \log (k) \log(q) / \eps^2)
    \]
    coordinates. Because the ultimate result from calling our sparsification procedure is the \emph{union} of all of the leaves of the recursive tree, the returned result has size at most 
    \[
    2^{\log \log k} \cdot O(k \eta \log (k) \log(q) / \eps^2) = O(k \eta \log^2(k)\log(q) / \eps^2),
    \]
    with probability at least $1 - 2^{-(0.19\eta - 110) \log k} \cdot k^{-100}$. 

    Finally, note that we can replace $\eps$ with a value $\eps' = \eps / 2 \log \log k$. Thus, the resulting sparsifier will be a $(1 \pm \eps')^{\log \log k} \leq (1 \pm \eps)$ sparsifier, with the same high probability.

    Taking the union bound of our errors, we can conclude that with probability $1 - 2^{-(0.19\eta - 110) \log k} \cdot k^{-100}$, Algorithm \ref{alg:CodeSparsify} returns a $(1 \pm \eps)$ sparsifier for $\calC$ that has at most $O(k \eta \log^2(k) \log(q) (\log \log (k))^2 / \eps^2)$ coordinates.
\end{proof}

\begin{theorem}\label{thm:codeSparsifyGeneralLength}
    For a code $\calC$ of dimension $k$, and length $n$ over $\F_q$, Algorithm \ref{alg:CodeSparsify} creates a $(1 \pm \eps)$ sparsifier for $\calC$ with probability $1 - \log (n) \cdot 2^{-(0.19\eta - 110) \log k} \cdot k^{-100}$ with at most \[
    O(k \eta \log(k) \log(q) \log^2(n) (\log \log (n))^2 / \eps^2)
    \]
    coordinates.
\end{theorem}

\begin{proof}
    For a code of dimension $k$, and length $n$, this means that our value of $d$ as specified in the first call to Algorithm \ref{alg:CodeSparsify} is at most $n$ as well. As a result, after only $\log \log n$ iterations, $d = n^{1 / 2^{\log \log n}} = n^{1 / \log n} = O(1)$. So, by Corollary \ref{clm:overallCodeAccuracy}, because the maximum recursion depth is only $\log \log n$, it follows that with probability at least $1 - (2^{\log \log n} -1) \cdot 2^{-(0.19\eta - 110) \log k} \cdot k^{-101}$, the returned result from Algorithm \ref{alg:CodeSparsify} is a $(1 \pm \eps)^{\log \log n}$ sparsifier for $\calC$.

    Now, by Lemma $\ref{clm:overallSpace}$, with probability $\geq 1 - 2^{\log \log n} \cdot 2^{-\eta k} \geq 1 - \log (n) \cdot 2^{-(0.19\eta - 110) \log k} \cdot  2^{-k}$, every code at recursive depth $\log \log n$ has at most 
    \[
    (1 + 1 / 2 \log \log (k))^{\log\log n} \cdot n^{1/\log n} \cdot \eta \cdot k \log(k)\log(q) / \eps^2 = O(k \eta \log (k) \log(q) \cdot e^{\frac{\log \log n}{\log \log k}} / \eps^2)
    \]
    coordinates. Because the ultimate result from calling our sparsification procedure is the \emph{union} of all of the leaves of the recursive tree, the returned result has size at most 
    \[
    \log (n)  \cdot e^{\frac{\log \log n}{\log \log k}} \cdot O(k \eta \log (k)\log(q) / \eps^2) = O(k \eta \log(k) \log(q) \log^2(n) / \eps^2),
    \]
    with probability at least $1 - \log (n) \cdot 2^{-(0.19\eta - 110) \log k} \cdot k^{-101}$. 

    Finally, note that we can replace $\eps$ with a value $\eps' = \eps / 2 \log \log n$. Thus, the resulting sparsifier will be a $(1 \pm \eps')^{\log \log n} \leq (1 \pm \eps)$ sparsifier, with the same high probability.

    Taking the union bound of our errors, we can conclude that with probability $1 - \log (n) \cdot 2^{-(0.19\eta - 110) \log k} \cdot k^{-100}$, Algorithm \ref{alg:CodeSparsify} returns a $(1 \pm \eps)$ sparsifier for $\calC$ that has at most 
    \[
    O(k \eta \log(k)\log(q) \log^2(n) (\log \log (n))^2 / \eps^2)
    \]
    coordinates.
\end{proof}

However, as we will address in the next section, this result is not perfect: 
\begin{enumerate}
    \item For large enough $n$, there is no guarantee that this probability is $\geq 0$ unless $\eta$ depends on $n$.
    \item For large enough $n$, $\log^2(n)$ may even be larger than $k$.
\end{enumerate}

\section{Nearly Linear Size Sparsifiers for Codes of Arbitrary Length}\label[customsection]{sec:anylengthsparse}

In this section, our goal is to prove \Cref{thm:main} (the exact version proved will be Theorem \ref{thm:mainRestated}). We will do this by using Theorem \ref{thm:codeSparsifyGeneralLength} as a sub-routine in another algorithm.

In the previous section, we saw an algorithm which produces a near-linear size sparsifier for codes of length polynomial in the dimension. However, if we start with a code of arbitrary length $n$, simply applying the algorithm from the previous section led to spurious $\log(n)$ factors, which unfortunately can dwarf $k$. In this section, we will show how we can be a little more careful with our sparsifier to avoid these extra $\log(n)$ factors. To do this, in \Cref{sec:quadratic}, we will showcase a simple one-shot algorithm that returns a size $O(k^2 \log(q) / \eps^2)$ weighted $(1 \pm \eps)$ sparsifier of any code of dimension $k$ and length $n$. Ideally, we could simply use this sparsification and compose on top of it Algorithm \ref{alg:CodeSparsify}. However, as written, Algorithm \ref{alg:CodeSparsify} only works for \emph{unweighted} codes (or codes where every coordinate has the same weight).

Further, the ratio of the weights that are returned by this sparsifier is unbounded in $k$ (and in many cases will be as large as $\Omega(n)$). Naive notions of turning the weighted code into an unweighted code unfortunately do not work, as if we try to replace coordinates of weight $w$ with $w$ unweighted coordinates, we will no longer be guaranteed that length of the code is polynomial, and we will not have gained anything.

Instead, we will show that once we have a weighted sparsifier of size polynomial in the dimension, we can group coordinates together by their weights. That is, we set a parameter $\alpha = \text{poly}(k / \eps)$ sufficiently large, and set the $i$th group to contain coordinates with weights between $[\alpha^{i-1}, \alpha^i]$. Our key observation is that if a codeword is non-zero in any coordinate in the $i$th group, then for an appropriately chosen $\alpha$, the total weighted contribution from any coordinates in groups $i-2, i-3, \dots$ is much less than an $\eps / 100$ fraction of the weight coming from group $i$. Thus if we let $i$ be the largest integer such that a codeword is non-zero in the $i$th weight group, we can effectively ignore all the coordinates corresponding to weight groups $i-2, \dots$ when sparsifying this codeword. Now, starting with the largest $i$, we decompose the code into codewords which are non-zero in group $i$ and those which are zero in group $i$. For those which are non-zero, we can effectively ignore all the coordinates from groups $i-2, i-3, \dots$. This means that all the coordinates we are concerned with have weights in the range $[\alpha^{i-2}, \alpha^i]$. To turn this into an unweighted code, we simply pull out a factor of $\alpha^{i-2}$, and now for a coordinate of weight $w$, we can repeat it roughly $w$ times. Because the weights are polynomial in the dimension, the resulting unweighted code is also polynomial in dimension, and we can invoke the results from the previous section. 

\subsection{Simple Quadratic Size Sparsifiers}\label[customsection]{sec:quadratic}

In this section, we will introduce a one-shot method for sparsifying a code of dimension $k$ and length $n$ on alphabet $\F_q$ that maintains $O(k^2 \log(q) / \eps^2)$ indices of the original code. We state the algorithm here, and then analyze the space complexity and correctness of this algorithm.

\begin{algorithm}\label{alg:SimpleQuadratic}
    \caption{QuadraticSparsify$(\calC \subseteq \F_q^n, k, \eps)$}
    Let $n$ be the length of $\calC$. 
    \For{$i = 1 , \dots n$}{
    Let $w_i$ be $\min_{c \in \calC: c_i \neq 0} \wt(c)$.
    }
    Let $\calC'$ be the result of sampling every coordinate of $\calC$ with probability $\min(1, a \cdot k \log(q) / (\eps^2 w_i))$, and weight $1 / \min(1, a \cdot k / (\eps^2 w_i))$. \\
    \Return{$\calC'$}
\end{algorithm}

\subsubsection{Correctness}

First, we prove the correctness of this algorithm. 
\begin{lemma}
    For a code $\calC$ of dimension $k$, and a fixed codeword $c \in \calC$, Algorithm \ref{alg:SimpleQuadratic} returns a sparsifier $\calC'$ for $\calC$, such that the new weight of $c$ is a $(1 \pm \eps)$ approximation to the old weight with probability at least $1 - 2^{-2k}$.
\end{lemma}
\begin{proof}
Consider any codeword $c \in \calC$ of weight $\ell$. Then, by Claim \ref{clm:concentrationBound},
\[
\Pr[\calC' \text{does not make a }(1\pm\eps)\text{ approximation to }c ] \leq 2e^{-0.38 \eps^2 \ell \frac{a \cdot k \log(q)}{\eps^2 \ell}} = 2e^{-0.38a k \log(q)}.
\]
Here, we have used that because $w_i$ is the minimum weight codeword for which a specific coordinate is $1$, in a codeword of weight $\ell,$ every index in the support has $w_i \leq \ell$. Now, by choosing $a = 10$, and taking a union bound over all $q^k$ codewords, we get that with probability $\geq 1 - 2^{-k}$, Algorithm \ref{alg:SimpleQuadratic} returns a $(1 \pm \eps)$ sparsifier for $\calC$.
\end{proof}

\subsubsection{Size Analysis}

Next, we bound the space taken by this algorithm. To do this, we will first need to take advantage of some structural results about codes.

\begin{fact}
    For any linear code, there exists a basis such that codewords of weight $\ell$ can be written as the sum of codewords of weight $\leq \ell$ from the basis. 
\end{fact}

\begin{proof}
Fix a basis $b_1, \dots b_k$, which maximizes the number of codewords which can be written as the sum of codewords of weight less than itself. Suppose that this basis cannot write some codeword $c$ as a sum of codewords of weight $\leq \wt(c)$. We know that $c = b_1 + \dots b_k$ for some basis elements. Without loss of generality, assume that the weights of $b_1, \dots b_k$ are all ordered. Further, assume that all basis elements starting at index $1 \leq j \leq k$ are of weight $\geq \wt(c)$. It follows then that we can swap $c$ with $b_k$. For any old codeword which required basis element $b_k$ in its decomposition, we can substitute $b_k = b_1 + \dots + b_{k-1} + c$. All of these basis elements are of weight $\leq \wt(b_k)$, so if previously the codeword could be written as the sum of basis elements of weight less than itself, this will still hold true. Further, the new code can express $c$ in its basis with codewords of weight $\leq \wt(c)$ (just by taking itself). Hence, the new basis is better than the original, which is a contradiction. So, the original basis must be able to write every codeword as the sum of codewords of smaller weight. 
\end{proof}

\begin{remark}\label{rmk:basis}
    Fix such a basis $b_1, \dots b_k$ as specified by the previous fact. Now, consider any coordinate of this code. If we look at the weights of all codewords that are non-zero in this coordinate, the minimum weight codeword must be with one of the basis vectors. This follows easily: any codeword non-zero in its $i$th coordinate must be the sum of at least $1$ basis vector which is non-zero in its $i$th coordinate. This means that the weight of the codeword must be greater than the weight of the corresponding basis vectors in its sum. 
\end{remark}

By the previous remark, when we try to bound the possible values attained by $\max_x \frac{\langle r_i, x \rangle}{\wt(\calD x)}$, it will suffice to analyze the possible value attained just by looking at the basis codewords for this special basis. This simplifies things, as we do not have to look at what can happen with possible linear combinations of the codewords. Instead of looking at a basis, we may simply consider a matrix of dimension $n \times k$. 

\begin{claim}\label{claim:boundbasis}
    Fix an $n \times k$ matrix $A$. Let $c_i$ denote the $i$th coordinate of $c$, and let $A_j$ denote the $j$th column of $A$. Then,
    \[
    \sum_{i = 1}^n \max_{j \in [k]: (A_j)_i = 1} \frac{1}{\wt(c_j)} \leq k.
    \]
\end{claim}
\begin{proof}
    This follows because each column $c_j$ can only be the \say{minimizing} column $\wt(c_j)$ times. Each such time, it contributes $\frac{1}{\wt(c_j)}$. There are $k$ columns, so the total contribution is thus bounded by $k$. This is in fact tight, as the identity matrix will achieve $k$.
\end{proof}

\begin{claim}\label{clm:boundSumOfRecipWeights}
    For any linear code $\calC$ of dimension $k$ and length $n$, with a generator matrix $G$ consisting of rows $r_i$, 
    \[
    \sum_{i = 1}^n\max_{c \in \calC: c_i = 1} \frac{1}{\wt(c)} \leq k.
    \]
\end{claim}

\begin{proof}
    This follows by taking the specified codeword basis from Remark \ref{rmk:basis} and invoking Claim \ref{claim:boundbasis}.
\end{proof}

Finally, we can prove a bound on the size of the sketch. 
\begin{lemma}\label{clm:simpleQuadraticSize}
    With probability $1 - 2^{-k}$, Algorithm \ref{alg:SimpleQuadratic} does not sample more than $O(k^2 \log(q) / \eps^2)$ coordinates.
\end{lemma}

The expected number of coordinates that are sampled by Algorithm \ref{alg:SimpleQuadratic} is 
\[
\sum_{i = 1}^n \min(1, a \cdot k\log(q) / (\eps^2 w_i)) \leq \sum_{i = 1}^n a \cdot k\log(q) / (\eps^2 w_i) \leq \frac{ak\log(q)}{\eps^2} \cdot \sum_{i = 1}^n 1 / w_i \leq \frac{ak^2\log(q)}{\eps^2}. 
\]

Note that here we have used Claim \ref{clm:boundSumOfRecipWeights}. Finally, by using a Chernoff bound, we can argue that the size of the sketch is not more than double its expected size with probability $1 - 2^{-k}$. Hence, the size of the sketch is at most $O(k^2\log(q) / \eps^2)$ with probability $1 - 2^{-k}$.

\subsection[Removing the O(log n) factors]{Removing the $O(\log n)$ factors}

Similar to \cite{CKN20}, we want to remove the extra factors of $\log n$. To this end, we suggest the following procedure upon being given a code of length $n$ and dimension $k$ in Algorithm \ref{alg:WeightClassDecomposition}. 

\begin{algorithm}\label{alg:WeightClassDecomposition}
\caption{WeightClassDecomposition$(\calC, \eps, k)$}
    Let $\calC' = $QuadraticSparsify$(\calC, k, \eps/4)$. \\
    Let $\alpha = \frac{k^3 \log(q)}{\eps^3}$. \\
    Let $E_i$ be all coordinates of $\calC'$ that have weight between $[\alpha^{i-1}, \alpha^i]$. \\
    Let $\Dodd = E_1 \cup E_3 \cup E_5 \cup \dots$, and let $\Deven = E_2 \cup E_4 \cup E_6 \cup \dots$. \\
    \Return{$\Dodd, \Deven$}.
\end{algorithm}

Next, we prove some facts about this algorithm.

\begin{lemma}\label{clm:PreserveApproxWeightDecomp}
    Consider a code $\calC$ of dimension $k$ and length $n$. Let 
    \[
    \Dodd, \Deven = \mathrm{WeightClassDecomposition}(\calC, \eps, k).
    \] To get a $(1 \pm \eps)$-sparsifier for $\calC$, it suffices to get a $(1 \pm \eps/4)$ sparsifier to each of $\Dodd, \Deven$.
\end{lemma}

\begin{proof}
    First, note that in Algorithm \ref{alg:WeightClassDecomposition}, we let $\calC' = $QuadraticSparsify$(\calC, t, \eps/4)$. From before, we know that this will return a $(1 \pm \eps/4)$ sparsifier $\calC'$ to $\calC$, of size $O(k^2 \log(q) / \eps^2)$ with probability $\geq 1 - 2^{-k-1}$. Now, the creation of $\Dodd, \Deven$ forms a \emph{vertical} decomposition of the code $\calC'$. Thus, by Claim \ref{clm:verticalDecomp}, if we have a $(1\pm \eps/4)$ sparsifier for each of $\Dodd, \Deven$, we have a $(1\pm \eps/4)$ sparsifier to $\calC'$, so by Claim \ref{clm:composingApproximations} we have a $(1 \pm \eps)$ approximation to $\calC$ (with probability $1 - 2^{k-1}).$
\end{proof}

Because of the previous claim, it is now our goal to create sparsifiers for $\Dodd, \Deven$. Without loss of generality, we will focus our attention only on $\Deven$, as the procedure for $\Dodd$ is exactly the same (and the proofs will be the same as well). At a high level, we will take advantage of the fact that 
\[
\Deven = E_2 \cup E_4 \cup \dots ,
\]
where each $E_i$ contains edges of weights $[\alpha^{i-1}, \alpha^i]$, for $\alpha = \frac{k^3 \log(q)}{\eps^3}$. Because the returned result from Quadratic sparsify has at most $O(k^2\log(q) / \eps^2)$ edges with high probability, whenever a codeword $c \in \calC'$ has a $1$ in a coordinate corresponding to $E_i$, we can effectively ignore all coordinates of lighter weights $E_{i-2}, E_{i-4}, \dots$. This is because any coordinate in $E_{\leq i-2}$ has weight at most a $\frac{\eps^3}{k^3 \log(q)}$ fraction of any single coordinate in $E_{i}$. Because there are at most $O(k^2 \log(q) / \eps^2)$ coordinates in $\calC'$, it follows that the total possible weight of all coordinates in $E_{\leq i-2}$ is still at most a $O(\eps / k)$ fraction of the weight of a single coordinate in $E_{i}$. Thus, we will argue that when we are creating a sparsifier for codewords that have a $1$ in a coordinate corresponding to some $E_i$, we will be able to effectively ignore all coordinates corresponding to $E_{\leq i-2}$. To argue this, we will first have to show how to decompose the code into blocks that are non-zero in coordinates in $E_i$. So, consider the following algorithm:

\begin{algorithm}\label{alg:SingleSpanDecomposition}
    \caption{SingleSpanDecomposition$(\Deven, \alpha, i)$}
    Let $E_i$ be all coordinates of $\Deven$ with weights between $\alpha^{i-1}$ and $\alpha^i$. \\
    Let $G$ be a generating matrix for $\Deven$. \\
    Let $k'$ be the rank of $G|_{E_i}$. \\
    Let $b_1, \dots b_{k'}$ be $k'$ linearly independent columns in $G|_{E_i}$. \\
    Permute the columns of $\Deven$ so $b_1, \dots b_{k'}$ become the first $k'$ columns of $G|_{E_i}$. \\
    Perform column operations on $G$ to cancel out all remaining columns in $G|_{E_i}$. \\
    \Return{$G|_{E_i}$, $G|_{\bar{E_i}}$, $k'$}
\end{algorithm}

\begin{claim}
    Line 6 in Algorithm \ref{alg:SingleSpanDecomposition} is always possible.
\end{claim}

\begin{proof}
    Because the rank of $G|_{E_i}$ is $k'$, and the first $k'$ columns are said to be linearly independent, it follows that there exists a sequence of column operations we can do to zero-out all the remaining $k - k'$ columns of $G|_{E_i}$. 
\end{proof}

\begin{claim}
    Line 6 in Algorithm \ref{alg:SingleSpanDecomposition} does not change the span of the overall generating matrix $G$. 
\end{claim}

\begin{proof}
    Because the first $k'$ columns remain untouched, and then are added to the remaining $k- k'$ columns, it follows that all these operations can be undone. Since the starting generator matrix $G$ was rank $k$, it follows that the new generating matrix is also rank $k$, so the span has not changed. 
\end{proof}

We now describe some structural properties of this decomposition:

\begin{claim}\label{clm:cleanDecomp}
    After a single iteration of Algorithm \ref{alg:SingleSpanDecomposition}, suppose the result of running the algorithm looks like
        \[
 G = 
 \begin{bmatrix}
A & 0 \\
B & C
\end{bmatrix},
\]
where $A$ corresponds to the first $k'$ columns of $G_{E_i}$, and $B, C$ are the remaining coordinates of the decomposition. Then, if a codeword $c \in \Deven$ is $0$ on all of the coordinates corresponding to $E_i$ in the above matrix, then $c$ lives entirely in the span of the final $k-k'$ columns.
\end{claim}

\begin{proof}
    Suppose to the contrary that $c$ is $0$ on all of the coordinates corresponding to $E_i$, but requires a non-zero linear combination including the first $k'$ coordinates. Then $c$ would be non-zero in the coordinates corresponding to $E_i$ because $A$ is a set of linearly independent columns for $E_i$. Thus, any non-zero linear combination including the columns of $A$ will be non-zero in $E_i$, so $c$ can not include any of the first $k'$ columns.
\end{proof}

Now, we can continue to decompose the code by repeating Algorithm \ref{alg:SingleSpanDecomposition} multiple times.

\begin{algorithm}\label{alg:SpanDecomposition}
    \caption{SpanDecomposition$(\Deven, \alpha)$}
    Let $\Deven' = \Deven$. \\
    Let $S = \{ \}$.\\
    \While{$\Deven'$ is not empty}{
    Let $k$ be the dimension of $\Deven'$.
    Let $i$ be the largest integer such that $E_i$ is non-empty in $\Deven'$. \\
    Let $G|_{E_i}, G|_{\bar{E_i}}, k' =$SingleSpanDecomposition$(\Deven', \alpha, i)$. \\
    Let $H_i$ be the first $k'$ columns of $G|_{E_i}$. \\
    Let $\Deven'$ be the span of the final $k - k'$ columns of $G|_{\bar{E_i}}$. \\
    Add $i$ to $S$.
    }
    \Return{$S$, $H_i$ for every $i \in S$}
\end{algorithm}

\begin{claim}\label{clm:conserveRank}
    Let $S, H_i$ be as returned by Algorithm \ref{alg:SpanDecomposition}. Then, $\sum_{i \in S}\mathrm{rank}(H_i) = \mathrm{rank}(\Deven)$.
\end{claim}

\begin{proof}
    This follows because in line 6 of Algorithm \ref{alg:SpanDecomposition} we set $H_i$ to be the first $k'$ columns of $G|_{E_i}$, and recurse on $\Deven'$ being the span of the remaining $k-k'$ columns. Hence, the total rank is conserved in every inner loop.
\end{proof}

\begin{lemma}\label{clm:OnlyApproxHi}
    Suppose we have a code of the form $\Deven$ created by Algorithm \ref{alg:WeightClassDecomposition}. Then, if we run Algorithm \ref{alg:SpanDecomposition} on $\Deven$, to get $S, H_i \forall i \in S$, it suffices to get a $(1 \pm \eps/2)$ sparsifier for each of the $H_i$ in order to get a $(1 \pm \eps)$ sparsifier for $\Deven$.
\end{lemma}

\begin{proof}
    Consider any codeword $c$ in the span of $\Deven$. Let $j$ be the largest integer such that $c$ is non-zero in the coordinates of $\Deven$ corresponding to $E_j$. 
    
    First, we will show that it suffices to approximate the weight of $c$ to $(1 \pm \eps/2)$ on the coordinates in $E_j$ to approximate its weight to $(1\pm \eps/2)$ overall. Indeed, this follows because any single coordinate in $E_j$ has more weight than all the combined coordinates of $E_{j-2}, E_{j-4}, \dots$. This is because there are only $O(k^2 \log(q) / \eps^2)$ coordinates in the code total, and by our choice of $\alpha$ in Algorithm \ref{alg:WeightClassDecomposition}, any coordinate in $E_j$ is at least $k^3 \log(q) / \eps^3$ of the fraction of the weight of a coordinate in $E_{j-2}, E_{j-4}, \dots$. So, any single coordinate in $E_j$ contributes $\Omega(k / \eps)$ more weight than all coordinates in $E_{j-2}, \dots$ combined. If we approximate the weight of $c$ in $E_j$ to a $(1 \pm \eps /2)$ fraction then, we are never overestimating the weight of $c$ in the code by more than $(1 + \eps/2)$, and we never underestimate by more than $(1 - \eps / 2)(1 - O(\eps/k) ) \geq (1 - \eps)$. Hence, this does indeed yield a $(1 \pm \eps)$ approximation to the weight of a codeword $c$. 

    Next, we must argue that by creating sparsifiers for each of the $H_i$, we are indeed approximating the weight of any codeword $c$ to a $(1 \pm \eps/2)$ fraction on the coordinates of $\Deven$ corresponding to $E_j$. To see why this is true, let us look at a single iteration of Algorithm \ref{alg:SingleSpanDecomposition}. If WLOG we assume the coordinates of $\Deven$ are sorted by weight, the result of running the algorithm looks like 
    \[
 \calD_{\text{even}} = 
 \begin{bmatrix}
A & 0 \\
B & C
\end{bmatrix}.
\]

In this case, $A$ is a dimension $k'$ code corresponding to the coordinates of $E_j$ in $\Deven$. We then disregard the matrix $B$, and iteratively decompose $C$ in the same manner. The key fact is from Claim \ref{clm:cleanDecomp}. Any codeword which is zero on $E_j$ will in fact live entirely in the span of the final $k-k'$ columns. Thus, it suffices to estimate the weight of $c$ on these final $k-k'$ columns, but because the coordinates of $E_j$ are $0$ in these columns, it in fact suffices to simply build a sparsifier for the matrix $C$ in the above decomposition. Thus, inductively, it suffices to continue decomposing $C$ in the above manner. To conclude, if for some $c$, $j$ is the largest integer for which $c$ is non-zero on $E_{j}$, then in each iteration when we decompose the generator into 
    \[
 G^{(i)} = 
 \begin{bmatrix}
A & 0 \\
B & C
\end{bmatrix},
\]
$c$ will continue living in the span of the bottom right matrix $C$, until we finally call Algorithm $\ref{alg:SingleSpanDecomposition}$ with parameter $j$. Then, we are indeed approximating $c$ on the coordinates of $\Deven$ corresponding to $E_{j}$, so our argument is complete.
\end{proof}

\paragraph{Dealing with Bounded Weights}

Let us consider any $H_i$ that is returned by Algorithm \ref{alg:SpanDecomposition}, when called with $\alpha = k^3 \log(q) / \eps^3$. By construction, $H_i$ will contain weights only in the range $[\alpha^{i-1}, \alpha^{i}]$ and will have at most $O(k^2 \log(q) / \eps^2)$ coordinates. In this subsection, we will show how we can turn $H_i$ into an unweighted code  with at most $O(k^5 \log^2(q)/\eps^6)$ coordinates. First, note however, that we can simply pull out a factor of $\alpha^{i-1}$, and treat the remaining graph as having weights in the range of $[1, \alpha]$. Because multiplicative approximation does not change under multiplication by a constant, this is valid. Formally, consider the following algorithm:

\begin{algorithm}\label{alg:MakeUnweighted}
    \caption{MakeUnweighted$(\calC, \alpha, i, \eps)$}
    Divide all edge weights in $\calC$ by $\alpha^{i-1}$. \\
    Make a new unweighted code $\calC'$ by duplicating every coordinate of $\calC$ $\lfloor 10 w(r) / \eps \rfloor$ times. \\
    \Return{$\calC, \alpha^{i-1} \cdot \eps / 10$}
\end{algorithm}

\begin{lemma}\label{clm:PreserveApproxUnweighted}
Consider a code $\calC$ with weights bounded in the range $[1, \alpha]$. To get a $(1 \pm \eps)$ sparsifier for $\calC$ it suffices to return a $(1 \pm \eps / 10)$ sparsifier for $\calC'$ weighted by $\eps / 10$, where $\calC'$ is the result of calling Algorithm \ref{alg:MakeUnweighted} on $\calC, \alpha, 1, \eps$.
\end{lemma}

\begin{proof}
    It suffices to show that $\calC'$ is $(1 \pm \eps/10)$ sparsifier for $\calC$, as our current claim will then follow by Claim \ref{clm:composingApproximations}. Now, to show that $\calC'$ is $(1 \pm \eps/10)$ sparsifier for $\calC$, we will use Claim \ref{clm:verticalDecomp}. Indeed, for every coordinate $r$ in $\calC$, consider the corresponding $\lfloor 10 w(r) / \eps \rfloor$ coordinates in $\calC'$. We will show that the contribution from these coordinates in $\calC'$, when weighted by $\eps/10$, is a $(1 \pm \eps/10)$ approximation to the contribution from $r$. 

    So, consider an arbitrary coordinate $r$, and let its weight be $w$. Then,
    \[
    \frac{10w}{\eps} - 1 \leq \lfloor 10 w / \eps \rfloor \leq \frac{10w}{\eps}.
    \]

    When we normalize by $\frac{\eps}{10}$, we get that the combined weight of the new coordinates $w'$ satisfies 
    \[
    w - \eps/10 \leq w' \leq w.
    \]

    Because $w \geq 1$, it follows that this yields a $(1 \pm \eps/10)$ sparsifier, and we can conclude our statement.
\end{proof}

\begin{claim}\label{clm:boundUnweightedSupport}
    Suppose a code $\calC$ of length $n$ has weight ratio bounded by $\alpha$, and minimum weight $\alpha^{i-1}$. Then, calling Algorithm \ref{alg:MakeUnweighted} with error parameter $\eps$ yields a new unweighted code of length $O(n \alpha / \eps)$.
\end{claim}

\begin{proof}
    Each coordinate is repeated at most $O(\alpha / \eps)$ times.
\end{proof}

\subsection{Final Algorithm}

Finally, we state our final algorithm in Algorithm \ref{alg:FinalCodeSparsify}, which will create a $(1 \pm \eps)$ sparsifier for any code $\calC \subseteq \F_q^n$ of dimension $k$ preserving only $\widetilde{O}(k \log(q) / \eps^2)$ coordinates.

\begin{algorithm}\label{alg:FinalCodeSparsify}
    \caption{FinalCodeSparsify$(\calC, \eps)$}
    Let $k$ be the dimension of $\calC$. \\
    Let $\alpha = k^3 \log(q) / (\eps/2)^3$, and $\Dodd, \Deven = $WeightClassDecomposition$(\calC, \eps, k)$. \\
    Let $S_{\text{even}}, \{ H_{\text{even}, i} \} = $SpanDecomposition$(\Deven, \alpha)$. \\
    Let $S_{\text{odd}}, \{ H_{\text{odd}, i} \} = $SpanDecomposition$(\Dodd, \alpha)$. \\
    \For{$i \in S_{\text{even}}$}{
    Let $\widehat{H}_{\text{even}, i}, w_{\text{even}, i}=$ MakeUnweighted$({H}_{\text{even}, i}, \alpha, i, \eps/8)$. \\
    Let $\widetilde{H}_{\text{even}, i} = $CodeSparsify$(\widehat{H}_{\text{even}, i}, \mathrm{rank}(\widehat{H}_{\text{even}, i}), \eps/80, 100 (\log(k/\eps) \log \log(q))^2)$.
    }
    \For{$i \in S_{\text{odd}}$}{
    Let $\widehat{H}_{\text{odd}, i}, w_{\text{odd}, i}=$ MakeUnweighted$({H}_{\text{odd}, i}, \alpha, i, \eps/8)$. \\
    Let $\widetilde{H}_{\text{odd}, i} = $CodeSparsify$(\widehat{H}_{\text{odd}, i}, \mathrm{rank}(\widehat{H}_{\text{odd}, i}), \eps/80, 100 (\log(k/\eps) \log \log(q))^2)$.
    }
    \Return{$\bigcup_{i \in S_{\text{even}}} \left ( w_{\text{even}, i} \cdot \widetilde{H}_{\text{even}, i} \right ) \cup \bigcup_{i \in S_{\text{odd}}} \left ( w_{\text{odd}, i} \cdot \widetilde{H}_{\text{odd}, i} \right )$}
\end{algorithm}

First, we analyze the space complexity. WLOG we will prove statements only with respect to $\Deven$, as the proofs will be identical for $\Dodd$.

\begin{claim}\label{clm:singleCallSparse}
    Suppose we are calling Algorithm \ref{alg:FinalCodeSparsify} on a code $\calC$ of dimension $k$. Let $\teveni = \mathrm{rank}(\widehat{H}_{\text{even}, i})$ from each call to the for loop in line 5. 
    
    For each call $\widetilde{H}_{\text{even}, i} = $CodeSparsify$(\widehat{H}_{\text{even}, i}, \mathrm{rank}(\widehat{H}_{\text{even}, i}), \eps/10, 100 (\log(k/\eps) \log \log(q))^2)$ in Algorithm \ref{alg:FinalCodeSparsify}, the resulting sparsifier has
    \[
    O\left ( \teveni \log(\teveni) \log^2(k/\eps) \cdot \log^2(k/\eps) \log(q) (\log \log (k/\eps) \log \log(q))^2 / \eps^2 \right )
    \]
    coordinates with probability at least $1 - \log(k\log(q) / \eps) \cdot 2^{-\Omega(\log^2(k / \eps) (\log \log (q))^2)}$.
\end{claim}

\begin{proof}
    We use several facts. First, we use Theorem \ref{thm:codeSparsifyGeneralLength}. Note that we have replaced the $n$ in the statement of Theorem \ref{thm:codeSparsifyGeneralLength} with $k^5 \log^2(q) / \eps^6$ by using Claim \ref{clm:boundUnweightedSupport}. Indeed, because $\alpha = k^3 \log(q) / \eps^3$, and we started with a weighted code of length $O(k^2 \log(q) / \eps^2)$, it follows that after using Algorithm \ref{alg:MakeUnweighted}, the support size is bounded by $O(k^5 \log^2(q) / \eps^6)$. We've also added the fact that $\eta$ is no longer a constant, and instead carries $O((\log(k/\eps) \log \log(q))^2)$, and carried this through to the probability bound. 
\end{proof}

\begin{lemma}\label{clm:overallDevenSize}
    In total, the combined number of coordinates over $i \in S_{\text{even}}$ of all of the $\widetilde{H}_{\text{even}, i}$ is at most $\widetilde{O}(k \log(q) / \eps^2)$ with probability at least $1 - \log(k\log(q) / \eps) \cdot 2^{-\Omega(\log^2(k / \eps) (\log \log (q))^2)}$.
\end{lemma}

\begin{proof}
    First, we use Claim \ref{clm:conserveRank} to see that 
    \[
    \sum_{i \in S_{\text{even}}} \teveni \leq k,
    \]
    where $\teveni = \mathrm{rank}(\widehat{H}_{\text{even}, i})$. 
    Thus, in total, the combined length (total number of coordinates preserved) of all the $\widetilde{H}_{\text{even}, i}$ is 
    \begin{align*}
        & \sum_{i \in S_{\text{even}}} \text{number of coordinates in }\widehat{H}_{\text{even}, i} \\
        &\leq \sum_{i \in S_{\text{even}}} O\left ( \teveni \log(\teveni) \log^2(k/\eps) \cdot \log^2(k/\eps) \log(q) (\log \log (k/\eps) \log \log(q))^2 / \eps^2 \right ) \\
        & \leq \sum_{i \in S_{\text{even}}} (\teveni) \cdot \widetilde{O} \left (\log^4(k)  \log(q) / \eps^2 \right ) \\
        & = k \cdot \widetilde{O} \left (\log^4(k)  \log(q) / \eps^2 \right ) \\
        & = \widetilde{O}(k\log(q) / \eps^2).
    \end{align*}

    To see the probability bound, we simply take the union bound over all at most $k$ distinct $\widetilde{H}_{\text{even}, i}$, and invoke Claim \ref{clm:singleCallSparse}.
\end{proof}

Now, we will prove that we also get a $(1 \pm \eps)$ sparsifier for $\Deven$ when we run Algorithm \ref{alg:FinalCodeSparsify}.

\begin{lemma}\label{clm:DevenCorrectness}
    After combining the $\widehat{H}_{\text{even}, i}$ from Lines 5-8 in Algorithm \ref{alg:FinalCodeSparsify}, the result is a $(1 \pm \eps/4)$-sparsifier for $\Deven$ with probability at least $1 - \log(k\log(q) / \eps) \cdot 2^{-\Omega(\log^2(k / \eps) (\log \log (q))^2)}$.
\end{lemma}

\begin{proof}
    
    We use Lemma \ref{clm:OnlyApproxHi}, which states that to sparsify $\Deven$ to a factor $(1 \pm \eps/4)$, it suffices to sparsify each of the $H_{\text{even}, i}$ to a factor $(1 \pm \eps / 8)$, and then combine the results.

    Then, we use Lemma \ref{clm:PreserveApproxUnweighted}, which states that to sparsify any $H_{\text{even}, i}$ to a factor $(1 \pm \eps / 8)$, it suffices to sparsify $\widehat{H}_{\text{even}, i}$ to a factor $(1 \pm \eps / 80)$, where again, $\widehat{H}_{\text{even}, i}$ is the result of calling Algorithm \ref{alg:MakeUnweighted}. Then, we must multiply $\widehat{H}_{\text{even}, i}$ by a factor $\alpha^{i-1} \cdot \eps / 10$.    
    
    Finally, the resulting code $\widehat{H}_{\text{even}, i}$ is now an unweighted code, whose length is bounded by $O(k^5 \log^2(q) / \eps^6)$, with rank $\teveni$. The accuracy of the sparsifier then follows from Theorem \ref{thm:codeSparsifyGeneralLength} called with parameter $\eps / 80$.

    The failure probability follows from noting that we take the union bound over at most $k$ $H_{\text{even}, i}$. By Theorem \ref{thm:codeSparsifyGeneralLength}, our choice of $\eta$, and the bound on the length of the support being $O(k^5 \log^2(q)/ \eps^5)$, the probability bound follows.
\end{proof}

The reason for our choice of $\eta$ is a little subtle. For Theorem \ref{thm:codeSparsifyGeneralLength}, the failure probability is characterized in terms of the dimension of the code that is being sparsified. However, when we call Algorithm \ref{alg:CodeSparsify} as a sub-routine in Algorithm \ref{alg:FinalCodeSparsify}, we have no guarantee that the rank is $\omega(1)$. Indeed, it is certainly possible that the decomposition in $H_i$ creates $k$ different matrices all of rank $1$. Then, choosing $\eta$ to only be a constant, as stated in Theorem \ref{thm:codeSparsifyGeneralLength}, the failure probability could be constant, and taking the union bound over $k$ choices, we might not get anything meaningful. To amend this, instead of treating $\eta$ as a constant in Algorithm \ref{alg:CodeSparsify}, we set $\eta = 100 (\log(k/\eps) \log \log(q))^2$, where now $k$ is the \emph{overall} rank of the code $\calC$, \emph{not} the rank of the current code that is being sparsified $H_i$. With this modification, we can then attain our desired probability bounds.

\begin{theorem}\label{thm:mainRestated}
For any code $\calC$ of dimension $k$ and length $n$, Algorithm \ref{alg:FinalCodeSparsify} returns a $(1 \pm \eps)$ sparsifier to $\calC$ with $\widetilde{O}(k \log(q) / \eps^2)$ coordinates with probability $\geq 1 - 2^{-\Omega((\log(k/\eps) \log \log(q))^2)} - 2^{-k}$.
\end{theorem}

\begin{proof}
First, we use Lemma \ref{clm:PreserveApproxWeightDecomp}. This Lemma states that in order to get a $(1 \pm \eps)$ sparsifier to a code $\calC$, it suffices to get a $(1 \pm \eps/4)$ sparsifier to each of $\Deven, \Dodd$, and then combine the results.

    Then, we invoke Lemma \ref{clm:DevenCorrectness} to conclude that with probability $\geq 1 - 2^{-\Omega((\log(k/\eps) \log \log(q))^2)}$, Algorithm \ref{alg:FinalCodeSparsify} will produce $(1 \pm \eps/4)$ sparsifiers for $\Deven, \Dodd$.

    Further, to argue the sparsity of the algorithm, we use Lemma \ref{clm:overallDevenSize}. This states that with probability $\geq 1 - 2^{-\Omega( (\log(k/\eps) \log \log(q))^2)}$, Algorithm \ref{alg:FinalCodeSparsify} will produce code sparsifiers of size $\widetilde{O}(k \log(q) / \eps^2)$ for $\Deven, \Dodd$. 

    Finally, we can bound the error probability of the subcall to WeightClassDecomposition in Algorithm \ref{alg:FinalCodeSparsify} by $2^{-k}$.

    Thus, in total, the failure probability is at most $2^{-k} + 2^{-\Omega((\log(k/\eps) \log \log(q))^2)}$, the total size of the returned code sparsifier is at most $\widetilde{O}(k \log(q) / \eps^2)$, and the returned code is indeed a $(1 \pm \eps)$ sparsifier for $\calC$, as we desire.

    Note that the returned sparsifier may have some duplicate coordinates because of Algorithm \ref{alg:MakeUnweighted}. Even when counting duplicates of the same coordinate separately, the size of the sparsifier will be at most $\widetilde{O}(k \log(q) / \eps^2)$. We can remove duplicates of coordinates by adding their weights to a single copy of the coordinate.
\end{proof}

\paragraph{Extension to the weighted case.}
Finally, note that as stated, this section proves the existence of near-linear size code sparsifiers for \emph{unweighted} codes of any length. However, this extends simply to \emph{weighted} codes of any length, as we can simply repeat coordinates in accordance with their weights, and then sparsify the resulting unweighted code.

We make this more rigorous below:

\begin{claim}
    Suppose we are given a weighted code $\calC$ of dimension $k$, where the weight of coordinate $i$ is $w_i \in \mathbb{R}^+$, and the smallest weight of any coordinate is $w$. Then, if we create a new unweighted code $\calC'$ by repeating coordinate $i$ $\lfloor \frac{100 w_i}{\eps w} \rfloor$ times, then for any $c \in C$, the corresponding $c' \in \calC'$ satisfies 
    \[
    \frac{w \eps}{100} \cdot \wt(c') \in (1 \pm \eps/10) \wt(c).
    \]
    Further, a $(1 \pm \eps/10)$-sparsifier to $\calC'$ when weighted by $\frac{w \eps}{100}$ yields a $(1 \pm \eps)$-sparsifier for $\calC$. Hence, there exist sparsifiers of size $\widetilde{O}(k \log(q) / \eps^2)$ for any weighted code $\calC$. 
\end{claim}

\begin{proof}
    We will consider an arbitrary coordinate $i$ from $\calC$ and compare its weight contribution to $c$ in $\calC$ versus $c' \in \calC'$. In $\calC$, the contribution is $w_i$. In $\calC'$ (after weighting the entire code by $\frac{w \eps}{100}$), the contribution from the duplicates of coordinate $i$ is $
    \frac{w \eps}{100} \cdot \lfloor \frac{100 w_i}{\eps w} \rfloor$. Now, it follows that 
    \[
     \frac{w \eps}{100} \cdot \left ( \frac{100 w_i}{\eps w}\right ) \leq \frac{w \eps}{100} \cdot \lfloor \frac{100 w_i}{\eps w} \rfloor \leq \frac{w \eps}{100} \cdot \left ( \frac{100 w_i}{\eps w} + 1 \right ),
    \]
    and hence 
    \[
    w_i \leq \frac{w \eps}{100} \cdot \lfloor \frac{100 w_i}{\eps w} \rfloor \leq w_i \cdot (1 + \frac{w \eps}{w_i 100} ) \leq w_i (1 + \eps / 100).
    \]

    Hence, (after weighting the entire code by $\frac{w \eps}{100}$), the contribution from the duplicates of coordinate $i$ is within $(1 \pm \eps/100)$ of the desired weight, and hence overall, the corresponding weight of $c'$ is preserved to a $(1 \pm \eps / 100)$.

    Now, suppose we get a $(1 \pm \eps / 10)$-sparsifier for $\calC'$. Then, this sparsifier preserves the weight of any codeword $c' \in \calC'$ to a $(1 \pm \eps / 10)$ factor. Hence, for a codeword $c \in \calC$, the sparsifier for $\calC'$ (when weighted by $w \eps / 100$) preserves the weight of $c$ to a factor $(1 \pm \eps)$ by Claim \ref{clm:composingApproximations} (composing approximations).

    The size of our resulting sparsifier is $\widetilde{O}(k \log(q) / \eps^2)$. This follows because duplicating coordinates does not change the dimension or our field size, only the length. Because after duplicating the coordinates, the code is unweighted, we can invoke Theorem \ref{thm:codeSparsifyGeneralLength} to create a $(1 \pm \eps/10)$-sparsifier of size $\widetilde{O}(k \log(q) / \eps^2)$. Note that this sparsifier may still have duplicate coordinates, which we can remedy by combining them together (adding their weights). 
\end{proof}

\section{Application to Cayley Graph Sparsifiers}\label[customsection]{sec:cayleyGraph}

In this section, we will explore the connection between the weights of codewords for an error correcting code $\calC$ of dimension $k$ and the eigenvalues of the Laplacian of a Cayley graph on $\F_2^k$. At a high level, it is well known that there exist Cayley graph expanders on $\F_2^k$ with constant expansion when the degree of the graph is $\Omega(k)$. Any such expander $H$ can be viewed as a sparsifier to the complete Cayley graph $G$ on $\F_2^k$ (where we take the set of generators $S$ to be exactly $\F_2^k$), under the constraint that the resulting sparsifier still has a Cayley graph structure, and $L_G \approx_{\eps} L_H$. At a high level, our result says that for \emph{any} Cayley graph $G$ over $\F_2^k$, there exists a sparsifier $H$ with at most $\widetilde{O}(k / \eps^2)$ edges, such that $H$ is still a Cayley graph, and $L_H \approx_{\eps} L_G$.

Note that by prior work \cite{BSS09}, we know that there exist sparsifiers $H$ for any Cayley graph $G$ such that $L_H \approx_{\eps} L_G$, however this is the first work which proves the existence of such graphs under the restriction that $H$ is also a Cayley graph and of nearly-linear size.  

\paragraph{Preliminaries}

First, we introduce some definitions related to Cayley graphs. In this section, we will fix a binary linear code $\calC$, as well as a generating matrix for $\calC$, denoted by $G_{\calC}$. Let $r_i$ denote the $i$th row of $G_{\calC}$.

\begin{Definition}
    A Cayley graph $G$ is a graph with algebraic structure; its vertex set is defined to be a group, and the edges correspond to a set of generators $S$, along with weight $(w_i)_{i \in S}$. For every element in $s \in S$, and for every vertex $v$, there is an edge from $v$ to $v + s$ of weight $w_s$.
\end{Definition}

\begin{Definition}
    Let $\chi_x(r) = (-1)^{\langle x, r \rangle}$, where the inner product is taken modulo $2$, and $x, r \in \F_2^k$.
\end{Definition}

\begin{fact}
    For a Cayley graph $G$ defined over $\F_2^k$ with generating set $S$, there is exactly one eigenvalue of $G$ for every vertex $x$ in its vertex set (which is $\F_2^k$). The corresponding eigenvalue (of the adjacency matrix) is 
    \[
    \lambda_x(G) = \sum_{r \in S} w_r \chi_x(r).
    \]
    The corresponding eigenvector is $\chi_x$. Note that this means that any Cayley graph defined on the same vertex set has the same eigenvectors.
\end{fact}

Going forward, we will let $G$ be a Cayley graph over $\F_2^k$, where its set of generators $S$ is exactly $\{ r_1, \dots r_n \}$, where these are the rows of the generating matrix $G_{\calC}$.

\begin{fact}
    For a message $x \in \F_2^k$, we have that 
    \[
    \mathrm{Bias}(G_{\calC} x) = \mathbb{E}_{r \in S} \chi_x(r_i).
    \]
\end{fact}

The above statement is very intuitive. $\chi_x(r_i)$ is $1$ if the $i$th bit in the codeword corresponding to $x$ is $0$, and is $-1$ if the $i$th bit is $1$. As a result, this expectation is exactly measuring how many more $0$'s there are than $1$'s.

\begin{fact}\label{fact:eigenvaluesCodewords}
    We can generalize the previous fact to the eigenvalues of the Laplacian. In this way we get that 
    \[
    \lambda_x(L_G) = n - \sum_{r \in S} \chi_x(r) = n (1 - \mathrm{Bias}(G_{\calC} x)) = 2 \cdot \wt(G_{\calC} x).
    \]
\end{fact}

\begin{claim}
    By preserving the weight of every codeword of $\calC$ to a $(1 \pm \eps)$ factor, we preserve the eigenvalues of $L_G$ to a $(1 \pm \eps)$ factor. 
\end{claim}

\begin{proof}
    This follows exactly from Fact \ref{fact:eigenvaluesCodewords}. If we have a code sparsifier $\hat{C}$ for $\calC$, then for any $x \in \F_2^k$, $\wt(G_{\hat{\calC}}) \in (1 \pm \eps) \wt(G_{\calC}x)$. Because the codeword weights of the generating set and the eigenvalues of the Cayley graph are exactly equal, this $(1 \pm \eps)$ approximation to the codeword weights implies that a Cayley graph with the same weighted generating set as used by the code sparsifier would be a $(1 \pm \eps)$ spectral sparsifier by Fact \ref{fact:eigenvaluesCodewords}.
\end{proof}

\section{Applications to Sparsifying CSPs}\label[customsection]{sec:csps}

In this section, we show how to use our result on the sparsifiability of codes in the setting of CSPs. Specifically we first show that all affine predicates are sparsifiable, thus proving \Cref{thm:affineCSPs}. Then we use this theorem to classify all Boolean ternary CSPs, CSPs on variables that take values in $\{0,1\}$ where the predicate applies on three variables. This leads to a proof of \Cref{thm:ternaryCSPs}. 

\subsection{Affine CSPs}

Recall that a predicate $P:\F_q^r \to \{0,1\}$ is an {\em affine} predicate if there exist elements $a_0,a_1,\ldots,a_r \in \F_q$ such that $P(b_1,\ldots,b_r)=0$ if and only if $a_0 + \sum_i a_i b_i = 0$ (over $\F_q$). We say further that $P$ is {\em linear} if $a_0 = 0$. 

The proof of \Cref{thm:affineCSPs} is completely straightforward if $P$ is linear, given the definition of a linear code. The extension to the affine case uses a simple reduction from the affine case to the linear case (with one extra variable). 

\begin{proof}[Proof of~\Cref{thm:affineCSPs}]
Given an instance $\Phi$ of CSP$(\calP)$ with variables $x_1,\ldots,x_k$ and constraints $C_1,\ldots,C_n$ where
$C_j = P^{(j)}(x_{(j), 1},\ldots,x_{(j), r_j})$, 
we will create a code $\calC \subseteq \F_q^n$ of dimension $k$ generated by the matrix $G \in \F_q^{n \times k}$, where each row of the generating matrix corresponds to a single constraint.
Let $P^{(j)} \in \calP$ denote the predicate showing up in the $j$th constraint of our CSP instance. We start with the case that $P^{(j)}$ is linear with elements $a_{(j), 1},\ldots,a_{(j), r_j} \in \F_q$ being such that $P^{(j)}(b_1,\ldots,b_{r_j})=0$ if and only if $\sum_i a_{(j), i} b_i = 0$. Then, in the corresponding $j$th row of the generating matrix, for each $i \in [r_j]$, we place $a_{(j), i}$ in the column corresponding to variable $x_{(j),i}$, and leave all other entries in the row to be $0$. 
It is straightforward to verify that for an assignment $x \in \F_q^k$, $(Gx)_j = 0$ if and only if $C_j$ is unsatisfied. Thus $\wt(Gx)$ counts the number of satisfied constraints for assignment $x$ and thus a code sparisifier for $\calC$ is a sparsifier for the instance $\Phi$ of CSP$(P)$.

Now considering the case of a general affine $P^{(j)}$ given by $P^{(j)}(b_1,\ldots,b_r)=0$ if and only if $a_{(j), 0} + \sum_i a_{(j), i} b_i = 0$. Now let $\widehat{P^{(j)}}(b_0,\ldots,b_r) = \sum_{i=0}^r a_{(j), i} b_i$. Note that $\widehat{P^{(j)}}$ is linear. Given an instance $\Phi$ of CSP$(\calP)$ on variables $x_1,\ldots,x_k$ with constraints $C_1,\ldots,C_n$ where $C_j = P^{(j)}(x_{(j), 1},\ldots,x_{(j), r_j})$, let $\hat{\Phi}$ be the instance of CSP$(\calP)$ on variables $x_0,\ldots,x_k$ with constraints $\hat{C}_1,\ldots,\hat{C}_n$ given by $\hat{C}_j = \widehat{P^{(j)}}(x_0,x_{(j), 1},\ldots,x_{(j), r})$. We note that an assignment $x \in \F_q^k$ for $\Phi$ corresponds to the assignment $(1,x) \in \F_q^{k+1}$ to $\hat{\Phi}$ and so a sparsifier for $\hat{\Phi}$ (available from the previous paragraph) also sparsifies $\Phi$. This concludes the proof.
\end{proof}

A major open question from the work of \cite{BZ20} was the sparsifiability of XOR predicates. Even on $3$ variables, it was not known if the predicate $P(x_1, x_2, x_3) = x_1 \oplus x_2 \oplus x_3$ was sparsifiable to near-linear size. As a consequence of \Cref{thm:affineCSPs}, we get the following result that resolves this question.

\begin{corollary}
    On a universe of $k$ variables, any CSP with $r$-XOR predicates for $1 \leq r \leq k$ is $(1 \pm \eps)$ sparsifiable to size $\widetilde{O}(k / \eps^2)$. 
\end{corollary}

\subsection{Ternary Boolean Predicates}

We now turn to the classification of ternary Boolean predicates. Recall that a predicate $P: \zo^r \ra \zo$ has {\em an affine projection to AND} if there exists a function $\pi:[r] \to \{0,1,x,\neg{x}, y, \neg{y}\}$ such that $\mathrm{AND}(x,y) = P(\pi(1),\ldots,\pi(r))$. We wish to prove that $P:\zo^3\to\zo$ is sparsifiable nearly linear size if and only if it has no affine projection to AND (\Cref{thm:ternaryCSPs}). 

The hardness result follows from a well-known result showing that the dicut problem is not sparsifiable to subquadratic size~\cite{FK17}, which in our language is equivalent to saying that the binary AND predicate is not sparsifiable. (We include a precise statement and proof below for completeness --- see~\Cref{lem:ANDhard}.)
Extending this to all predicates that have an affine projection to AND is simple (and holds for general $r$). This is stated and proved as \Cref{lem:affineANDhard} below. The bulk of the section then does a case analysis and shows that all ternary Boolean predicates that do not have an affine projection to AND can be sparsified by appealing to \Cref{thm:affineCSPs}. 

\paragraph{Non-sparsifiability of predicates with affine projection to AND.}

\begin{lemma}[\cite{FK17}]\label{lem:ANDhard}
    For every $\epsilon \in [0,1)$, every $(1 \pm \epsilon)$-sparsifier of size $s$ for CSP(AND) on $k$ variables requires $s = \Omega(k^2)$. 
\end{lemma}

The proof is actually more general and shows that any ``sketch'' of an instance $\Phi$ of CSP(AND) requires $\Omega(k^2)$ bits. 

\begin{proof}
    Let $S$ and $w_S$ be a $(1 \pm \epsilon)$ sparsifier of size $s$ of a CSP instance $\Phi$ on variables $x_1,\ldots,x_k$. Given a sparsifier, i.e., a subset of the constraints $S$ and a pair $(i,j) \in \binom{k}2$,
    consider the weight of the constraints satisfied by the assignment $x^{ij}$ given by $x^{ij}_k = 1$ if $k \in \{i,j\}$ and $0$ otherwise. This weight is positive in $\Phi$ if and only if the constraint $x_i \AND x_j$ appears in $\Phi$. Thus this weight is positive in the weighted sparsified instance using constraints from $S$ if and only if the constraint $x_i \AND x_j$ appears in $\Phi$ (since $\epsilon < 1$); and furthermore the weight is positive in the unweighted sparsfied instance on $S$ if and only if the constraint $x_i \AND x_j$ appears in $\Phi$. (The presence of the weights only affect the weight of the constraints that are satisfied, but not whether the number is positive or not.) Since this is true to every $(i,j) \in \binom{k}2$, it follows that $S$ allows us to reconstruct $\Omega(k^2)$ independent bits of information about $\Phi$ and thus by the pigeonhole principle $|S| \geq \binom{k}2 = \Omega(k^2)$ (for some instance $\Phi$). 
\end{proof}

\begin{lemma}\label{lem:affineANDhard}
    For every $r$, if a predicate $P: \zo^r \ra \zo$ has an affine projection to AND, then for every $\epsilon \in [0,1)$, every $(1 \pm \epsilon)$-sparsifier of size $s$ for CSP$(P)$ requires $s = \Omega_r(k^2)$. 
\end{lemma}

\begin{proof}
    Let $\pi:[r] \to \{0,1,x,\neg{x},y,\neg{y}\}$ be such that AND$(x,y) = P(\pi(1),\ldots,\pi(r))$. 
    Given an instance $\Phi$ of CSP(AND) on $k$ variables $x_1,\ldots,x_k$ we create an instance $\Psi$ of CSP$(P)$ on $2rk+2$ variables denoted
    $G_0,G_1, Y_{i,t,b}$ for $i\in[k], t\in[r],b \in \zo$ as follows: For every constraint AND$(x_i,x_j)$ in $\Phi$, we introduce the constraint
    $P(v_1, \dots, v_r)$ where for $t \in [r]$,
    
    \[
    v_t = \begin{cases}
    G_0 & \text{if } \pi(t)=0 \\
    G_1 & \text{if } \pi(t)=1 \\
    Y_{i,t,0} & \text{if } \pi(t) = x \\
    Y_{i,t,1} & \text{if } \pi(t) = \neg{x} \\
    Y_{j,t,0} & \text{if } \pi(t) = y \\
    Y_{j,t,1} & \text{if } \pi(t) = \neg{y}
    \end{cases}.
    \]
    Given an assignment to $x_1,\ldots,x_k$, it can be verified that the resulting predicate simulates AND$(x_i,x_j)$ if $G_0 = 0$, $G_1 = 1$, and $Y_{\ell,t,0} = x_\ell$ and $Y_{\ell,t,1} = \neg{x_\ell}$ for all $\ell \in [k]$ and $t\in [r]$. Thus a sparsification $(S,w_S)$ of $\Psi$ yields a sparsification of $\Phi$. From the lower bound in \Cref{lem:ANDhard}, we get that $s = \Omega(k^2)$. Relative to $k'$ the number of variables   of $\Psi$ we get that $s = \Omega((k'/r)^2)= \Omega_r(k'^2)$ as desired.
\end{proof}

\paragraph{Sparsifying 3-CSPs with no affine projections to AND.}

Finally we show that if a predicate $P: \zo^3 \ra \zo$ has no affine projection to AND, then there exists linear-size sparsifiers for $P$. We will make use of the following corollary of \Cref{thm:affineCSPs}:

\begin{corollary}\label{cor:sparseLinearEq}
    Suppose there exists a linear equation $E(x_1, x_2, x_3) = ax_1 + bx_2 + cx_3 + d \mod p$ (for $p$ a prime) such that the unsatisfying assignmments to a predicate $P(x_1, x_2, x_3): \zo^3 \ra \zo$ are exactly the assigments to $x_1, x_2, x_3$ such that $E$ evaluates to $0$, then a valued CSP containing this predicate can be sparsified to size $\widetilde{O}(k \log(p)  / \eps^2)$.
\end{corollary}

We note that the corollary is immediate from \Cref{thm:affineCSPs}, which even allows the variables to take values in all of $\mathbb{Z}_p$ while we only need them to take values in $\{0,1\}$. Restricting the set of assignments preserves sparsification and so the sparsifier from \Cref{thm:affineCSPs} certainly suffices to get \Cref{cor:sparseLinearEq}.

With this in hand, we will now use a case by case analysis to show how to write any predicate $P: \zo^3 \ra \zo$ with no affine projection to AND as a linear equation modulo some prime. We state four claims that cover the different cases, and prove them in turn. Given the four claims, and \Cref{lem:affineANDhard}, the proof of \Cref{thm:ternaryCSPs} is immediate. 

\begin{claim}\label{clm:easy}
    If $P:\zo^3\to\zo$ has zero, six, seven or eight satisfying assignments then $P$ is sparsifiable to nearly linear size. 
\end{claim}

\begin{claim}\label{clm:five}
    If $P:\zo^3 \to \zo$ has five satisfying assignments and $P$ has no affine projections to AND then $P$ is sparsfiable to nearly linear size.
\end{claim}

\begin{claim}\label{clm:four}
    If $P:\zo^3 \to \zo$ has four satisfying assignments and $P$ has no affine projections to AND then $P$ is sparsfiable to nearly linear size.
\end{claim}

\begin{claim}\label{clm:hard}
    If $P:\zo^3\to\zo$ has one, two or three satisfying assignments then $P$ has an affine projection to AND. 
\end{claim}

\begin{proof}[Proof of \Cref{thm:affineCSPs}]
If $P$ has an affine projection to AND, then by \Cref{lem:affineANDhard}, $P$ has no subquadratic sized sparsifiers. So assume $P$ has no affine projections to AND.  Then by \Cref{clm:hard} (in contrapositive form) $P$ has at least four satisfying assignments. And \Cref{clm:easy}-\Cref{clm:four} show that in all remaining cases $P$ is sparsifiable to nearly linear size.    
\end{proof}

Thus all that remains is to prove \Cref{clm:easy}-\Cref{clm:hard}. 
Before turning to the proofs of these claims we mention some basic symmetries that allows us to simplify all the cases. 
\begin{lemma}
    For every predicate $P:\zo^r\to\zo$, permutation $\pi:[r] \to [r]$ and index $b \in \{0,1\}^r$, let $P'(z_1,\ldots,z_r) = P(z_{\pi(1)} \oplus b_1,\ldots,z_{\pi(r)}\oplus b_r)$. Then $P$ has a nearly linear sparsifier if and only if $P'$ does. 
\end{lemma}

\begin{proof}
    The proof when $P'$ is just a permutation of the variables of $P$ (i.e., when $b = 0^r$) is straightforward - we map any constraint of an instance of CSP$(P)$ to the constraint obtained by permuting the sequence of variables according to $\pi$. It thus suffices to prove the lemma for the case where $b \ne 0^r$ and $\pi$ is the identity. (By composing the two steps we get the full lemma). For this case we use an idea similar to the idea in the proof of \Cref{lem:affineANDhard}. 
    
    We first note that we can assume w.l.o.g that variables of the instances to be sparsified come in $r$ blocks and each constraint application applies constraints in which the $t$th variable comes from the $t$th block. To see this suppose we have an instance $\Phi$ with variables $x_1,\ldots,x_k$ and suppose some constraint is $P(x_{i_1},\ldots,x_{i_r})$. We create a new instance $\Phi'$ on variables $x_{i,t}$ for $i \in [k]$ and $t \in [r]$ and replace the constraint above by the constraint $P(x_{i_1,1},\ldots,x_{i_r,r})$. We claim a sparsification of $\Phi'$ yields a sparsification of $\Phi$. (In $\Phi$ we are only interested in assignments in which the $r$ copies of a variable all take on the same value. The sparsification of $\Phi'$ yields an estimate of the number of satisfied constraints for all assignments including these). 

    Once the instances apply constraints to variables from distinct blocks, we can now negate any subset of variables of $P$. Fix $b \in \zo^r$ and let $P'(z) = P(z \oplus b)$.  Given a canonical instance $\Phi'$ of CSP$(P)$ as above, we can simply let $\hat{\Phi}$ be the CSP$(P')$ instance where every constraint application applies $P'$ instead of $P$ to the same sequence of variables. To compute the sparsification of $\Phi'$ we simply use a sparsification of $\hat{\Phi}$ and use the fact that the weight of constraints satisfied by an assignment $\{a_{i,t}\}_{i,t}$ in $\hat{\Phi}$ is the same as the weight of constraints satisfied by the assignment $\{a_{i,t}\oplus b_t\}_{i,t}$ in $\Phi'$. This allows us to use a sparsification of CSP$(P')$ to get a sparsification of CSP$(P)$ (and the other direction follows similarly). 
    
\end{proof}

We now proceed by cases. By \Cref{cor:sparseLinearEq}, in order to show near-linear size sparsifiability, it suffices to show that the unsatisfying assignments to these predicates can be written as solutions to a linear equation $\mod p$ for some prime $p$.

\begin{proof}[Proof of \Cref{clm:easy}]
\begin{enumerate}
    \item $P$ has 0 or 8 Satisfying assignments: in both cases, the predicate is a constant function. So, we can in fact sparsify to a single constraint (just a single constraint with weight equal to the number of total constraints). 
    \item $P$ has 7 Satisfying assignments: if there are 7 satisfying assignments to $P$, then this is simply an OR on $3$ variables. By the reduction provided in \cite{KK15}, along with known hypergraph sparsification results \cite{CKN20}, this can be sparsified to size $\widetilde{O}(k / \eps^2)$. 
    This can also be shown using \Cref{cor:sparseLinearEq}: W.l.o.g. $P$ is unsatisfied by the all zeros assignment. So $P(x,y,z) = 0$ if and only if $x+y+z = 0 \mod 5$ which fits within the framework of \Cref{cor:sparseLinearEq}.  
    \item 6 Satisfying assignments to $P$: if there are $6$ satisfying assignments to $P$, then by replacing $x_i$ with $\neg x_i$, we can assume WLOG that one of the unsatisfying assignments is $000$, as argued earlier. There are then $3$ cases:
    \begin{enumerate}
        \item The other unsatisfying assigmnent is at distance $3$ from $000$, i.e. $111$ is the other unsatisfying assignment. Then, these are exactly the unsatisfying assignments to $x_1 + x_2 + x_3 \mod 3$.
        \item The other unsatisfying assignment is at distance $2$ from $000$. Then, by our argument before, we can permute the bits of this other unsatisfying assignment, i.e. up to re-ordering $011$ is the other unsatisfying assignment. Then, these are exactly the unsatisfying assignments to $x_1 + 2x_2 + 3x_3 \mod 5$.
        \item The other unsatisfying assignment is at distance $1$ from $000$. Then, up to re-ordering / negation, the other unsatisfying assignment is $001$. Then, our expression is exactly $x_1 \vee x_2$, which is known to be sparsifiable by \cite{FK17}, or similarly, by viewing it as the equation $x_1 + x_2 \mod 3$.
    \end{enumerate}
\end{enumerate}
This concludes the proof of \Cref{clm:easy}.
\end{proof}

\begin{proof}[Proof of \Cref{clm:five}]
We consider $P$ that has five satisfying assignments. Again, we break into several cases. By replacing variables with their negations, and possibly reordering the variables, we can always assume one unsatisfying assignment is $000$. We consider some cases on the distance to the nearest other satisfying assignment.
\begin{enumerate}        
        \item Suppose one other unsatisfying assignment is at distance $1$ from $000$. WLOG, let this other unsatisfying assignment be $001$. Note that this means the final unsatisfying assignment must start with $11$, as otherwise it will contain an AND of arity $2$ (to see this, if we assume WLOG for the sake of contradiction that the first bit is a $0$, consider the affine projection where $x_1 = 0$, there are $3$ unsatisfying assignments remaining under this projection, and hence an AND).
        Further, by negating the third variable, we can get either $110$, or $111$ without changing the other two unsatisfying assignments. Thus, we assume the final unsatisfying assignment is $111$. Indeed, the only case we have to deal with here is when the predicate $P$ has unsatisfying assignments which are $000, 001, 111$. However, note that under the affine restriction where $x_1 = x_2$, this is exactly an AND, as the unsatisfying assignments will be $00, 01, 11$. Hence, sparsifying this expression can require $\Omega(k^2)$ constraints.
        
        \item Suppose one other unsatisfying assignment is at distance $2$ from $000$. WLOG let this other unsatisfying assignment be $011$. Note that if we included an unsatisfying assignment that was at distance $1$ from $011$, then by variable negation and re-ordering, we would be back in the previous case. Hence, the only other case which has not yet been considered is when the other unsatisfying assignment is also at distance $2$ from $011$ and $000$. WLOG let this other unsatisfying assignment be $110$. Then, note that we can express this as the zeros to $x_1 + x_2 + 2x_3 \mod 3$.
\end{enumerate}
\end{proof}

\begin{proof}[Proof of \Cref{clm:four}]
    Let $P$ have 4 satisfying assignments. Note then on the 3-dimensional hypercube, every face must have either $0, 2,$ or $4$ satisfying assignments (as otherwise $P$ has an affine projection to AND). If any of the faces has all $4$ satisfying assignments, then $P$ is exactly just $x_i$ or $\neg x_i$ (which is easy to sparsify).  So, suppose every face has $2$ satisfying assignments. Without loss of generality, suppose one of the satisfying assignments is $111$. Then, there are two cases:
    \begin{enumerate}
        \item There is a satisfying assignment at distance $1$ from $111$, WLOG let this be $110$. Then the other two satisfying assignments must be $000$ and $001$, as otherwise, there exists a face with more than $2$ satisfying assignments. If this is the case, then constraint can be expressed as $x_1 + x_2 + 1 \mod 2$.
        \item There is no satisfying assignment at distance $1$ from $111$. This means there are only satisfying assignments at distance $2$ from $111$, WLOG let this be $100$. Note that if $000$ is also a satisfying assignment in this case, then by negating all the variables, we are back in the previous case. Hence, the only other case is when $000$ is not a satisfying assignment, which means that all the satisfying assignments are $100, 001, 010, 111$, which is exactly $x_1 + x_2 + x_3 \mod 2$.
    \end{enumerate}   
\end{proof} 

\begin{proof}[Proof of \Cref{clm:hard}]
We now consider the case where $P$ has one, two or three satisfying assignments and prove that in each case it has an affine projection to AND. 
\begin{enumerate}
    \item 3 Satisfying assignments: Suppose the satisfying assignments are $a_1 a_2 a_3$, $b_1b_2b_3$ and $c_1 c_2 c_3$. Choose a coordinate such that not all the strings are equal on this coordinate. This means that by restricting this coordinate, there are either $1$ or $2$ satisfying assignments on the face. So, fix the coordinate to make $1$ satisfying assignment, which yields an AND.
    \item 2 Satisfying assignments: Suppose the satisfying assignments are $a_1 a_2 a_3$ and $b_1b_2b_3$. Choose a coordinate such that not all the strings are equal on this coordinate. By restricting this coordinate, we create a predicate with one satisfying assignment, and hence an AND.
    \item 1 Satisfying assignment: Let the satisfying assignment be $a_1 a_2 a_3$. Restrict $x_1 = a_1$. This yields a predicate with $1$ satisfying assignment, and hence an AND. 
\end{enumerate}
\end{proof}

This concludes the proof of \Cref{clm:easy}-\Cref{clm:hard} and thus the proof of \Cref{thm:ternaryCSPs}.

\section{Application to Hypergraph Cut Sparsifiers}\label[customsection]{sec:hypergraph}

In this section, we will show how our result implies the existence of near-linear size hypergraph cut sparsifiers. First, we introduce the definition of a hypergraph. 

\begin{Definition}
    A \emph{hypergraph} $G = (V, E)$ is a set of n vertices $V$, along with a set of hyperedges $E$. Each hyperedge $e$ is a subset of $V$, of any size.
\end{Definition}

Next, we introduce the definition of a \emph{cut} in a hypergraph.

\begin{Definition}
    For a hypergraph $G = (V, E)$, a cut in the hypergraph is a non-empty subset $S \subset V$. The size of the cut $S$ is the number of hyperedges in $E$ that are not completely contained in $S$ or $V - S$. Intuitively, this is the number of edges that cross between $S, V-S$. We use $\delta_S(G)$ to denote the crossing edges corresponding to $S$ in $G$.
\end{Definition}

With this, we can then state a consequence of our main result in the setting of hypergraphs. 

\begin{corollary}
    For a hypergraph $G = (V, E)$ on $k$ vertices, there exists a weighted sub-hypergraph $\hat{G}$ of $G$ with $\widetilde{O}(k / \eps^2)$ hyperedges, such that for any subset $S \subseteq V$,
    \[
    (1 - \eps) \wt(\delta_G(S)) \leq \wt(\delta_{\hat{G}}(S)) \leq (1 + \eps) \wt(\delta_G(S)).
    \]
\end{corollary}

At a high level, our proof takes advantage of the fact that we showed the existence of code sparsifiers over \emph{any} arbitrary field $\F_q$. In particular, for a hypergraph on $k$ vertices, we will choose a prime $q$ between $k$ and $2k$. Then, we will create a generating matrix for a code over $\F_q$ where each row of the generating matrix corresponds to a hyperedge in the hypergraph. To start, we create a generating matrix with $k$ columns. Now, for any hyperedge by $e$, we denote its size by $|e|$. If we analyze the row of the generating matrix corresponding to edge $e$, we then place a $1$ in the columns corresponding to vertices $e_1, \dots e_{|e|-1}$. For $e_{|e|}$ (i.e. the final vertex contained in the hyperedge), we place the value $q - |e| + 1$. In doing so, the row-sum of any row of the generating matrix will be exactly $0$. Indeed, for any $\zo$ weighted linear combination of the columns of this generating matrix, a row is identically $0$ if either \emph{all} of the vertices corresponding to the hyperedge are included in the linear combination, or none of them are. This is exactly the definition of a hypergraph cut.

First, we use a basic number theoretic result:

\begin{fact}[Bertrand's Postulate]\label{fact:bertrand}
For any positive integer $n$, there exists a prime between $n$ and $2n$.
\end{fact}

Thus, for any hypergraph on $k$ vertices, we can find a prime number between $k, 2k$. Next, we define more specifically the generating matrix corresponding to a hypergraph:

\begin{Definition}
    For a hypergraph $H = (V, E)$ on $k$ vertices, let $q$ be a prime between $k, 2k$ as guaranteed by \Cref{fact:bertrand}. Let $G$ be a generating matrix of a code defined over $\F_q^{|E|}$, and let $G$ have $k$ columns. Now, for any hyperedge $e = v_1, \dots v_{|e|}$, let $G_{e, v_i} $ be $1$ if $i \leq |e| - 1$, and $q - |e| +1$ if $i = |e|$. All other entries in the row are zero. Call this generating matrix $G$ the \emph{generating matrix associated with $H$}.
\end{Definition}

\begin{remark}\label{rmk:hypergraphEncoding}
    Let $G$ be the generating matrix associated with a hypergraph $H$. Let $S\subseteq[k]$, and let $x$ be the indicator vector for $S$. Then,  
    \[
    \wt(\delta_H(S)) = \wt(Gx).
    \]
\end{remark}

\begin{proof}
    Consider any such $S \subseteq [k]$. $\wt(\delta_H(S))$ is exactly the number of hyperedges that are not completely contained in $S$ or $V-S$. Now, $\wt(Gx)$ is the number of non-zero entries in $Gx$. By our construction of $G$, the only way for a $\zo$ row-sum of $G$ to be zero is when a $\zo$ vector assigns either all $0$'s, or all $1$'s to the corresponding vertices of the hyperedge. This is exactly the same as the hyperedge being completely contained in $S$ or $V-S$.
\end{proof}

\begin{claim}
    Let $G$ be the generating matrix associated with a hypergraph $H$. If, there exists a sparsifier $\hat{G}$ such that for every message $x \in \F_q^k$, $\wt(\hat{G}x) \in (1 \pm \eps) \wt(Gx)$, then if we select the corresponding edges of $H$ with the same weights as in $\hat{G}$, we will recover a $(1 \pm \eps)$ hypergraph cut sparsifier for $H$.
\end{claim}

\begin{proof}
    By the previous Remark, for any set $S \subseteq [k]$, and $x$ being the indicator vector for $S$,
    \[
    \wt(\delta_H(S)) = \wt(Gx).
    \]
    Further, given $\hat{G}$, we can create the corresponding hypergraph $\hat{H}$. It is still true that 
    \[
    \wt(\delta_{\hat{H}}(S)) = \wt(Gx).
    \]
    Thus, we conclude that for any $S, x = \mathbf{1}[S]$,
    \[
    (1 - \eps ) \wt(\delta_H(S)) = (1 - \eps)\wt(Gx) \leq \wt(\hat{G}x) = \wt(\delta_{\hat{H}}(S)) \leq  (1 + \eps)\wt(Gx) = (1 + \eps ) \wt(\delta_H(S)).
    \]
    It follows that the hypergraph associated with $\hat{G}$ is indeed a hypergraph cut-sparsifier for $H$. 
\end{proof}

\begin{corollary}
    For any hypergraph $H$ on $k$ vertices, there exists a hypergraph cut sparsifier of $H$ with $\widetilde{O}( k / \eps^2)$ weighted hyperedges.
\end{corollary}

\begin{proof}
    Let $G$ be the generating matrix associated with $H$. Let $\hat{G}$ be the $(1 \pm \eps)$ code-sparsifier for the code generated by $G$. Note that because $q \leq 2k$, the number of rows in $\hat{G}$ is 
    \[
    \widetilde{O}(k \log q / \eps^2) = \widetilde{O}(k / \eps^2).
    \]
    By the previous claim, we can then let $\hat{H}$ be the hypergraph associated with $\hat{G}$. 
\end{proof}

Finally, by using Corollary \ref{cor:newKarger}, we can state a novel fact about the decomposition of hypergraphs.

\begin{corollary}\label{cor:hypergraphDecomp}
    For any hypergraph $H$ on $n$ vertices, for any integer $d \geq 1$, there exists a set of at most $nd$ hyperedges, such that upon their removal, the resulting hypergraph has at most $(2n)^{2\alpha}$ cuts of size $\leq \alpha d$.
\end{corollary}

\begin{proof}
    Let $G$ be the generating matrix associated with $H$. By the previous corollary, it follows that there exists a prime $q \leq 2n$, and a set of at most $nd$ rows we can remove from $G$ such that the number of codewords of weight $\leq \alpha d$ is at most $q^{\alpha} \cdot \binom{k}{\alpha}$. However, each such codeword corresponds with a possible cut in the graph of size at most $\alpha d$. Hence, the number of possible cuts in the graph of size $\leq \alpha d$ is at most $(2n)^{2\alpha}$.

    Note that again, if two separate vertex cuts $S_1, V - S_1$ and $S_2, V - S_2$ lead to exactly the same hyperedges being cut, we do not consider these to be separate cuts. Indeed, when we bound the number of cuts, we are bounding the number of distinct sets of hyperedges being cut of a given size.
\end{proof}

\section{Conclusions}

In this work, we showed that for any linear code $\calC \subseteq \F_q^n$ of dimension $k$, there exists a weighted set $S$ of $\widetilde{O}(k \log (q) / \eps^2)$ coordinates, such that for any codeword $c \in \calC$, the quantity $\wt(c|_S)$ is a $(1 \pm \eps)$ approximation to $\wt(c)$. This result provides a unified approach to recover known results about existence of near-linear size graph  and hypergraph cut-sparsifiers, as well as some new results that include near-linear size Cayley-graph sparsifiers of Cayley graphs over $\F_2^k$, and near-linear size sparsifiers for a broader class of CSPs than were previously known. 

The existential nature of our sparsification result raises the following natural question. Is there a poly-time algorithm to find the decomposition of Theorem \ref{lem:newKarger}? Alternately, is there a poly-time algorithm to compute code sparsifiers of near-linear size? The NP-hardness of the Minimum Distance Problem (MDP) for linear codes makes it particularly hard to implement Algorithm \ref{alg:SimpleQuadratic} (which explicitly calculates the minimum distance) and likewise finding the decomposition of Theorem \ref{lem:newKarger} for arbitrary $d$ can in \emph{some} cases solve the MDP as well. 

Another interesting direction for future work is to extend our classification theorem for sparsifiability of CSPs to predicates of arity greater than $3$.

\section*{Acknowledgments}

We thank Salil Vadhan for pointing out the connection between codes and the eigenvalues of the Laplacians of Cayley graphs over $\F_2^k$.

\bibliographystyle{alpha}
\bibliography{ref}

\appendix

\section{A Simpler Construction of Cut Sparsifiers for Graphs}\label[customsection]{sec:simplerBK}

Let $G = (V, E)$ be a graph on $n$ vertices, and $\eps \in (0,1)$ be some given accuracy parameter. Our goal in this section is to output a $(1 \pm \eps)$ cut-sparsifier for $G$. The first proof showing the existence of cut-sparsifiers of near-linear size was given by~\cite{BK96}, using the notion of a strength decomposition of a graph. Subsequently, a different proof based on edge connectivities was given in~\cite{FHH11}. While the starting point for these proofs is Karger's cut counting bound~\cite{Kar93, Kar99}, they both require additional ideas and are somewhat involved. We present here a simpler proof that directly utilizes Karger's cut counting bound to recursively sparsify $G$ while preserving its cuts.

At a high level, the algorithm is very simple: starting with a graph $G$ that has $d \cdot n \log (n) / \eps^2$ edges, we first remove all edges involved in cuts in this graph that have size $\leq \sqrt{d} \log(n) / \eps^2$, and place these edges in a graph $G_1$. By a simple argument, we can show that $G_1$ will have at most $n \sqrt{d} \log(n) / \eps^2$ edges. Now, the resulting graph from removing all the small cuts from $G$ is a graph we call $G'$. $G'$ has the special property that all non-zero cuts in $G'$ have size at least $\sqrt{d} \log(n) / \eps^2$. So, we can decompose $G'$ into its connected components, and using Karger's cut-counting bound for each of these connected components, we can argue that sampling each edge with probability $1 / \sqrt{d}$ and weight $\sqrt{d}$ will preserve cut-sizes with high probability. As a result, with high probability $G'$ has $\sqrt{d} n\log(n) / \eps^2$ edges, and cut-sizes are preserved. Thus, starting with a graph on $d \cdot n \log (n) / \eps^2$, we get two graphs on $\sqrt{d} \cdot n \log (n) / \eps^2$ edges. By repeating this procedure recursively, we can then get near-linear size cut-sparsifiers. 

Note that, as stated, the algorithm only works for unweighted graphs, and indeed our proof here is intended only for unweighted graphs in the name of simplicity. However, this proof can be extended to weighted graphs by using the more advanced decomposition techniques from \Cref{sec:anylengthsparse}.

In this section, the algorithm we propose will create cut sparsifiers with 
\[
O \left (\frac{n ( \log (n) \log \log (n))^2}{\eps^2} \right )
\] edges. The only non-trivial fact we will use is Theorem \ref{thm:kargerCutCounting}. Our algorithm is presented in Algorithm \ref{alg:GraphSparsify}.

\newcommand{\GraphSparsify}{\mathrm{GraphSparsify}}

\begin{algorithm}\label{alg:GraphSparsify}
\caption{$\GraphSparsify(G, \eps, n, i)$}
\If{$i = \log \log n$ or $G$ has $\leq n \log (n) / \eps^2$ edges}{
\Return $G$
}
Initialize empty graphs $G_1, G_2$ on $n$ vertices.

\While{$\exists$ a non-empty cut $(S, V - S)$ in $G$ of size $\leq \gamma(n)\cdot n^{1/2^{i}}$}{
Remove the edges in $\delta_G(S)$ from $G$, and place them in $G_1$.
}
\For{the remaining edges $e \in G$}{
Sample each $e$ with probability $\frac{1}{n^{1/2^{i}}}$, and if sampled, add it to $G_2$.
}
\Return $\GraphSparsify(G_1, \eps, n, i+1) \cup n^{1/2^{i}} \cdot \GraphSparsify(G_2, \eps, n, i+1)$.
\end{algorithm}

Note that in the algorithm, we are using $\gamma(n) = C \cdot \log n$, where $C$ is a constant (depending on $\eps$). We will need some claims in order to prove this result.

\begin{claim}
    For any $n$-vertex graph $G(V,E)$ and a positive integer $c \in [1..(n-1)]$, to remove all non-empty cuts of size $\leq c$ in the graph, we require removing only $(n-1)c$ edges.
\end{claim}
\begin{proof}
Fix a $c \in [1..(n-1)]$. Let $T(n)$ denote the maximum number of edges that are involved in cuts of size at most $c$ in an $n$-vertex graph.

    We prove this claim by induction on number of vertices. For our base case, consider any graph on $p \le (c+1)$ vertices. There are at most $\binom{p}{2} \leq \frac{(c+1)(p-1)}{2} \le (p-1) \cdot c$ edges in this graph, and all of them are in cuts of size at most $c$, so $T(p) \le (p-1)c$. Now, consider an arbitrary $n$-vertex graph. We will repeatedly remove from $G$ edges in cuts of size at most $c$. Upon finding a cut $(S, V-S)$ of size at most $c$, we remove all edges involved in the cut, and recursively continue on $G[S]$ and $G[V - S]$. So assuming $|S| = s$, we get that 
    \[
    T(n) \leq c + T(s) + T(n-s),
    \]
    where $1 \leq s \leq n-1$. Invoking the inductive hypothesis, we get $T(n) \leq c + (s-1)c + (n-s -1)c = (n-1)c$, completing the proof.
\end{proof}

\begin{claim}
    At the $i$th level of recursion, with high probability, each non-empty $G_i$ has at most $(1 + \frac{1}{\log \log n})^i \cdot n^{1 + 1/2^i} \cdot \gamma(n)$ edges, where $\gamma(n) = C \log n$. 
\end{claim}

\begin{proof}
We prove the claim by induction. Consider the first level of recursion of the algorithm. Let $n^{1/2} \cdot \gamma(n)$ be the minimum cut value on which we set our threshold. Then, the number of edges that we keep corresponding to the minimum cuts is at most $n^{3/2} \cdot \gamma(n)$. So, the first of the graphs in the subcall is of size at most $\gamma(n) \cdot n^{3/2}$. 

Now, we sample the remaining edges with probability $\frac{1}{n^{1/2}}$. Because the support size is $\geq n \log(n) / \eps^2$, we can use a Chernoff bound to argue that with high probability, at most $(1 + 1 / \log \log n)$ the expected number of edges will be included by the sampling procedure. This means that at most \[
n^2 \cdot \frac{(1 + 1 / \log \log n)}{n^{1/2}} = (1 + 1 / \log \log n)n^{3/2} \leq (1 + 1 / \log \log n) \cdot \gamma(n) \cdot n^{3/2}
\]
edges will be included with high probability. So, the number of edges in the second graph on which we recurse is at most $(1 + 1 / \log \log n) \gamma(n) \cdot n^{3/2}$.

Now, we suppose the claim holds by induction. Let us analyze the $j$th level of recursion. So, let the graph at this level be $G$, and have at most $(1 + 1 / \log \log n)^j \cdot n^{1 + 1/2^j} \cdot \gamma(n)$ edges. At this level of recursion, the first subgraph we make has all cuts of size at most $\gamma(n) n^{1/2^{j+1}}$. By our previous claim, there are at most $n \cdot \gamma(n) n^{1/2^{j+1}}$ many edges involved in those cuts, so our first graph we recurse on will satisfy the desired bound.
To construct the second graph we recurse on, we sample all remaining edges with probability $\frac{1}{n^{1/2^{j+1}}}$. By Chernoff again, we know the number of edges sampled is with very high probability at most $(1 + 1 / \log \log n)$ the expected number, because the graph has at least $n \log(n) / \eps^2$ edges. Hence, with high probability, the size of the second graph we recurse on has at most 
\[
\frac{(1 + 1 / \log \log n)}{n^{1/2^{j+1}}} \cdot (1 + 1 / \log \log n)^j \cdot n^{1 + 1/2^j} \cdot \gamma(n) = \gamma(n) \cdot (1 + 1 / \log \log n)^{j+1} n^{1 + 1/2^{j+1}}
\]
edges.
\end{proof}

\begin{claim}
    Suppose a graph $G$ is decomposed into two graphs $G = G_1 \cup G_2$, and we are given a $(1 \pm \eps)$ cut-sparsifier $H_1$ for $G_1$, $H_2$ for $G_2$. Then, $H_1 \cup H_2$ provides a $(1 \pm \eps)$ approximation to every cut in $G$.
\end{claim}

\begin{proof}
    Consider any cut $S, V- S$ in $G$. The claim follows because
    \begin{align*}
    (1 - \eps) \wt(\delta_G(S)) &= (1 - \eps)\wt(\delta_{G_1}(S)) + (1 - \eps) \wt(\delta_{G_2}(S)) \leq \wt(\delta_{H_1}(S)) + \wt(\delta_{H_2}(S)) \\
    &\leq (1 + \eps)\wt(\delta_{G_1}(S)) + (1 + \eps) \wt(\delta_{G_2}(S)) = (1 + \eps)\wt(\delta_G(S)).
    \end{align*}
\end{proof}

\begin{claim}
    If a cut-sparsifier $G''$ is a $(1 \pm \eps)$ cut-approximation to $G'$, and $G'$ is a $(1 \pm \delta)$ cut-approximation to $G$, then $G''$ is a $(1-\delta)(1-\eps), (1+\delta)(1 + \eps)$ approximation to $G$.
\end{claim}

\begin{proof}
    Consider any cut $(S, V-S)$. We know that $(1 - \eps)\wt(\delta_{G''}(S)) \leq \wt(\delta_{G'}(S)) \leq (1 + \eps)\wt(\delta_{G''}(S))$. Additionally, $(1 - \delta)\wt(\delta_{G'}(S)) \leq \wt(\delta_{G}(S)) \leq (1 + \delta)\wt(\delta_{G''}(S))$. Composing these two facts, we get our claim. 
\end{proof}

\begin{lemma}
    For any $j \in [1, \dots \log \log n]$, the output of our algorithm when called with $i = \log \log n - j$ on a graph $G$ is a $(1 \pm \eps)^j$ cut-sparsifier for $G$ with probability at least $1 - 4^{j}/n^8$.
\end{lemma}

\begin{proof}
    In the base case, we consider when $j = 0$. Clearly then, we return $G$, which is indeed a $1$-approximation. 
    
    Now, suppose the claim holds by induction. Now, let $i = \log \log (n) - j$. The algorithm, after receiving $G$, breaks $G$ in $G_1, G_2$, where $G_1$ contains all edges that are in cuts of size $\leq \gamma(n) \cdot n^{1 / {2^i}}$, and $G_2$ contains all remaining edges. Note that from our previous claims, it suffices to argue that we get $(1 \pm \eps)$ approximations to $G_1, G_2$, as the returned sparsifier for $G_1, G_2$ will be $(1\pm \eps)^{j-1}$ approximations by induction.
    
    Our algorithm completely preserves $G_1$, so this is not an issue. Instead, we focus on $G_2$. The algorithm samples every edge from $G_2$ with probability $\frac{1}{n^{1/2^{i+1}}}$. Because the minimum cut size in each component of $G_2$ is $\geq \gamma(n) n^{1/2^i}$, we know that the number of cuts in the graph of size $\leq \alpha \cdot \gamma(n) \cdot n^{1/2^i}$ is at most $n^{2 \alpha}$. This is because if we denote the sizes of the components as $x_1, \dots x_r$, $\sum_{i = 1}^r x_i^{2\alpha} \leq n^{2\alpha}$. Now, if we preserve every cut in each component of size $\leq \alpha c$, it follows that we preserve every cut in $G_2$ of size $\leq \alpha c$ (since the empty cuts are preserved trivially). For a cut of size $[\alpha/2 \gamma(n) \cdot n^{1/2^i}, \alpha \gamma(n) \cdot n^{1/2^i}]$, if we sample with probability $\frac{1}{n^{1/2^{i+1}}}$, then the probability that we do not get a $(1 \pm \eps)$ approximation to the cut is at most
    \[
    2 \cdot 2^{-0.38 \eps^2 \frac{\alpha}{2} \gamma(n) \cdot n^{1/2^i} \cdot n^{-1/2^{i+1}}} = 2 \cdot 2^{-0.19\eps^2 \alpha \gamma(n) \cdot n^{1/2^{i+1}}}.
    \]
    Taking the union bound over the at most $n^{2 \alpha}$ cuts of this size, the probability that we fail for cuts of size between $[\alpha/2, \alpha] \cdot C \cdot n^{1/2^i}$ is at most 
    \[
    2 \cdot 2^{-0.19 \alpha \gamma(n) \eps^2 \cdot n^{1/2^{i+1}}} \cdot 2^{2 \alpha \log n} \leq 2^{\alpha (-0.19 \eps^2 C \cdot n^{1/2^{i+1}} \log n + 3 \log n)}.
    \]

    Setting $C = \frac{100}{\eps^2}$, we get that with probability at most $1/n^{10}$ the sparsifier for cuts of size $[\alpha/2, \alpha] \cdot \gamma(n) \cdot n^{1/2^{i}}$ will fail. Now, we can take a union bound over $\alpha = 1, 2, \dots n^2$ to conclude that \emph{all} cuts in $G_2$ will be preserved to a $(1 \pm \eps)$ fraction with probability $ 1-1/n^8$. 

    Now, by induction, each of the recursive calls will return cut sparsifiers with probability $1 - 4^{j-1}/n^8$. So, the total failure probability at level $j$ is bounded by $4^{j-1}/n^8 + 4^{j-1}/n^8 + 1/n^8 \leq 4^j/n^8$, as we desire.

    Continuing to $j = \log \log n$, we get our desired result. 
\end{proof}

\begin{lemma}
    After recursing to depth $\log \log n$, the final returned graph has at most \[
    O(n \log^2 (n) (\log \log (n))^2 / \eps^2)
    \] edges with high probability. 
\end{lemma}

\begin{proof}
    We set $\eps' = \frac{\eps}{2 \log \log n}$. Then, the error in approximation for our starting graph is $(1-\frac{\eps}{2 \log \log n})^{\log log n}, (1 +\frac{\eps}{2 \log \log n})^{\log \log n}$, which yields a $(1 \pm \eps)$ cut-sparsifier with high probability. 
    
    At the $\log \log n$th level of recursion, each graph is of size at most 
    \[
    (1 + 1 / \log \log n)^{\log \log n} \cdot n \cdot n^{1 / 2^{\log \log n}} \cdot \gamma(n) = O(n \cdot \log (n) (\log \log (n))^2 / \eps^2),
    \]
    where we use that $\gamma(n) = O(\frac{ \log n}{\eps^2})$.
    Now, we take the union bound over all $2^{\log \log n}$ graphs at level $\log \log n$ in the recursion to conclude that there are at most $O(n \cdot \log^2 (n) (\log \log (n))^2 / \eps^2)$ edges in the sparsified graph with high probability. 
\end{proof}

\end{document}